\pdfoutput=1
\documentclass[conference]{IEEEtran}
\IEEEoverridecommandlockouts
\usepackage{cite}
\usepackage{amsmath,amssymb,amsfonts,amsthm}
\usepackage{graphicx}
\usepackage{textcomp}
\usepackage{xcolor}
\usepackage{bbding}
\usepackage{algorithm}
\usepackage{algorithmicx}
\usepackage{algpseudocode}
\usepackage[mathscr]{eucal}
\usepackage{multirow}
\usepackage{setspace}
\usepackage{color,framed}
\usepackage{xcolor}
\usepackage{mathrsfs}
\usepackage{tabularx}
\usepackage{utfsym}
\usepackage{soul}
\usepackage{subcaption}
\usepackage{float}
\usepackage{url}
\usepackage{enumitem}
\usepackage{balance}
\usepackage{booktabs}
\usepackage{threeparttable}
\usepackage[colorlinks,bookmarksopen,bookmarksnumbered,citecolor=green]{hyperref}
\usepackage{pifont}
\usepackage{wasysym}
\usepackage[defaultcolor=red]{changes}

\usepackage[most]{tcolorbox}
\definecolor{keys}{RGB}{0,135,0}
\definecolor{gray}{RGB}{216,216,216}
\def\BibTeX{{\rm B\kern-.05em{\sc i\kern-.025em b}\kern-.08em
    T\kern-.1667em\lower.7ex\hbox{E}\kern-.125emX}}

\newcommand{\code}[1]{{\fontfamily{cmtt}\fontseries{m}\fontshape{n}\selectfont\small{#1}}}

\usepackage{xspace}
\newcommand{\sysname}{\textsc{ODDFuzz}\xspace}
\newcommand{\refappendix}[1]{\hyperref[#1]{Appendix~\ref*{#1}}}

\begin{document}

\title{\huge \sysname: Discovering Java Deserialization Vulnerabilities via Structure-Aware Directed Greybox Fuzzing
%\thanks{$^\ast$Work done during internship at Ant Group.}
%\thanks{\textsuperscript{\Envelope}Corresponding author.}
}

\author{
    \IEEEauthorblockN{Sicong Cao$^{\dag\ast}$, Biao He$^\ddag$, Xiaobing Sun$^\dag$\textsuperscript{\Envelope}, Yu Ouyang$^\ddag$, Chao Zhang$^\S$, Xiaoxue Wu$^\dag$, Ting Su$^\P$, \\ 
    Lili Bo$^\dag$, Bin Li$^\dag$, Chuanlei Ma$^\ddag$, Jiajia Li$^\ddag$, Tao Wei$^\ddag$}
    \IEEEauthorblockA{$^\dag$Yangzhou University\quad $^\ddag$Ant Group\quad $^\S$Tsinghua University\quad $^\P$East China Normal University}
    \IEEEauthorblockA{$^\dag$\{DX120210088, xbsun, xiaoxuewu, lilibo, lb\}@yzu.edu.cn,}
    \IEEEauthorblockA{$^\ddag$\{hb187361, yu.oyy, chuanlei.mchl, jiajia.lijj, lenx.wei\}@antgroup.com,}
    \IEEEauthorblockA{$^\S$chaoz@tsinghua.edu.cn, $^\P$tsu@sei.ecnu.edu.cn}
}

\maketitle

\begin{abstract}
Java deserialization vulnerability is a severe threat in practice. Researchers have proposed static analysis solutions to locate candidate vulnerabilities and fuzzing solutions to generate proof-of-concept (PoC) serialized objects to trigger them.
However, existing solutions have limited effectiveness and efficiency.

% mechanism has been shown prone to cause security vulnerabilities. Being able to construct gadget chains is one of the key aspects of successfully exploiting insecure deserialization. Fuzzing is an effective testing technique to discover vulnerabilities. The main challenge of fuzzing deserialization gadget chains is to generate syntactically and semantically valid inputs such that security-sensitive sinks can be triggered. However, failure to handle the sophisticated nested structures of injection objects will make the fuzzing process stuck in the initial fuzzing stage, resulting in inefficient fuzzing.

In this paper, we propose a novel hybrid solution \sysname to efficiently discover Java deserialization vulnerabilities.
First, \sysname performs lightweight static taint analysis to identify candidate {\em gadget chains} that may cause deserialization vulnerabilities.
In this step, \sysname tries to locate all candidates and avoid false negatives.
Then, \sysname performs directed greybox fuzzing (DGF) to explore those candidates and generate PoC testcases to mitigate false positives.
Specifically, \sysname applies a structure-aware seed generation method to guarantee the validity of the testcases, and adopts a novel hybrid feedback and a step-forward strategy to guide the directed fuzzing.

We implemented a prototype of \sysname and evaluated it 
on the  popular Java deserialization repository \emph{ysoserial}. Results show that, \sysname could discover 16 out of 34 known gadget chains, while two state-of-the-art baselines only identify three of them. 
In addition, we evaluated \sysname on real-world applications including \code{Oracle WebLogic Server}, \code{Apache Dubbo}, \code{Sonatype Nexus}, and \code{protostuff}, and found six previously unreported exploitable gadget chains with five CVEs assigned.

% To address the above challenges,  we propose C{\scriptsize HASER}, a novel fuzzing framework for the deserialization gadget chain within Java applications. First, C{\scriptsize HASER} performs lightweight taint analysis to identify potentially suspicious gadgets chains. Then, C{\scriptsize HASER} applies a structure-aware seed generation to guarantee the validity of the constructed injection objects. Furthermore, C{\scriptsize HASER}  combines step-forward mutation with hybrid feedback to speed up the fuzzing efficiency. 
% We   implemented C{\scriptsize HASER} and evaluated it on the  popular Java deserialization gadget chain repository \emph{ysoserial}. Evaluation results show that C{\scriptsize HASER} can efficiently hunt 16 of 34 known gadget chains without false positives, including 13 unique gadget chains missed by state-of-the-art techniques. We also  carried out the gadget chain hunting in WebLogic, Sonatype Nexus, Apache Dubbo, and protostuff with C{\scriptsize HASER}. In total, we discovered six previously unreported exploitable gadget chains with five CVEs assigned.
\end{abstract}

\iffalse
\begin{IEEEkeywords}
Open dynamic deserialization (ODD), Java Deserialization Vulnerabilities, directed greybox fuzzing, Web security
\end{IEEEkeywords}
\fi

\section{Introduction}\label{Introduction}

The serialization mechanism \cite{DBLP:journals/toplas/HerlihyL82}, which is supported by mainstream programming languages like Java, JavaScript, PHP, and .NET, enables an application to convert an object to a stream of bytes for cross-process or cross-platform data transmission and persistence storage \cite{SALSA}. The counterpart of serialization is deserialization, which reconstructs an object from a serialized byte stream. 
This deserialization process is \textit{dynamic}, as different objects lead to polymorphic runtime behaviors.
Advanced language features (e.g., Java reflection \cite{Reflection1}) make the process even more dynamic.
% facilitating the inter-process communication and improving the performance of modern systems.
This process is also \textit{open}, i.e., crafted serialized objects may be injected by adversaries, which breaks the traditional trust boundary of inter-process data transmission and introduces attack surfaces.

Applications that unsafely deserialize incoming serialized objects would be abused to invoke a series of methods on the classpath, named \emph{gadget chains}, and eventually hijack security-sensitive code (e.g., {\code{Method.invoke()}}) or cause other consequences (e.g., access control bypass) \cite{DBLP:journals/smr/WeiSBCXL21,DBLP:journals/chinaf/SunPZLC19}.
% or denial of service).
% ot {\tt Runtime.exec()}).
Such \textit{open dynamic deserialization} (ODD) vulnerabilities are prevalent and  devastating.
The past few years have seen a proliferation of deserialization attacks in famous Java applications. For example, a recent zero-day vulnerability (named \code{Spring4Shell} \cite{Spring4Shell}) discovered in the \code{Spring} Framework \cite{Spring} allows an attacker to send a specially crafted HTTP request to bypass protections in the library's HTTP request parser, leading to remote code execution (RCE). Due to the dominance of \code{Spring} framework in the Java ecosystem, a large number of applications could potentially be impacted.

% While the prevalence of deserialization facilitates the inter-process communication and improves the performance of modern systems, this dynamic feature may pose security problems and has attracted the attention of attackers. \textcolor{blue}{From the attacker’s perspective, deserializing data from any other provenance allows attackers to inject crafted objects to abuse the application logic. Such an incoming serialized injection object could potentially conceal instructions that will force the target program to invoke a series of methods on its classpath, named \emph{gadget chains}, to perform well-designed payloads.} 

% Considering that the impact of deserialization vulnerabilities can be devastating, 
A limited number of tools~\cite{ysoserial,marshalsec,joogle,Scanner,Inspector,Serhybrid} have been proposed to discover ODD vulnerabilities in Java applications. 
% exploitable gadget chains, i.e., 
% From adversaries' perspective, 
The root cause of ODD vulnerabilities is that, the deserialized objects can \textit{reach} (in terms of control flow) and \textit{affect} (in terms of data flow) the sensitive code (sinks) of target applications.
Therefore, a straightforward way to discover ODD vulnerabilities is static taint analysis, as GadgetInspector \cite{Inspector} does.
However, such a purely static solution may suffer precision issues due to the limited support for Java deserialization-related features \cite{SALSA,sound}, resulting in both high false-negative and high false-positive rates. Furthermore, it requires manual inspection of the reports, which is \emph{time-consuming} and \emph{error-prone}. To alleviate this problem, SerHybrid~\cite{Serhybrid} adopts a hybrid analysis solution, which analyzes the heap access paths to find source objects that affect security-sensitive call sites, and utilizes fuzzing~\cite{fuzz,fuzzn} to generate source injection objects to verify whether the sinks are reachable.

% Two state-of-the-art and representative  solutions are GadgetInspector \cite{Inspector} and SerHybrid \cite{Serhybrid}. 
% GadgetInspector is the first automated gadget chain scanner, which utilizes static taint analysis to compute the propagation paths of parameters within/between methods of a target application to hunt gadget chains that can be exploited to execute deserialization attacks. 
% Since static analysis may suffer precision issues (i.e., a high false-positive rate) \cite{DBLP:conf/icse/UttureLKP22}, the developer has to manually examine and confirm the results, which is a labor-intensive process that requires a significant amount of expertise and usually error-prone. 
% To alleviate this problem, SerHybrid designs a hybrid analysis framework to statically hunt suspicious gadget chains and generate actual injection objects based on fuzzing for validation. \textcolor{blue}{Although SerHybrid confirms a few exploitable gadget chains without false positives, its effectiveness in gadget chain hunting is limited because a large amount of potential gadget chains are missed. A main reason lies in that statically identified gadget chains fail to be validated in dynamic fuzzing phase by randomly generated injection objects. As a result, existing approaches cannot effectively hunt Java deserialization gadget chains.}

However, these solutions in general have limited effectiveness and efficiency due to three challenges.
First, existing static analysis solutions struggle to make trade-offs between precision and recall. Due to the runtime polymorphism of Java language, any available overridden method (gadget) on the application's classpath may be exploited to construct gadget chains. Given that blindly enumerating all possible gadget chains will inevitably suffer from the path explosion problem, existing solutions often employ taint analysis \cite{FlowDroid} to prune infeasible gadget chains. However, they either are prone to precision issues \cite{Inspector}, or may not work due to the huge computation space caused by the prohibitive number of candidate gadget chains \cite{FUGIO}.
Second, existing fuzzing solutions are \textit{ineffective} at generating testcases (i.e., injection objects) to reach sinks. Note that, the injection objects may have a multilevel class hierarchy and their properties should satisfy certain control-flow or data-flow constraints. Fuzzing solutions without prior knowledge about such a complex nested form of structures are ineffective at generating qualified objects. 
Third, existing fuzzing solutions are \textit{inefficient} at generating testcases to reach sinks. They are \textit{coverage-guided} (i.e., trying to cover more code) rather than \textit{target-directed} (i.e., trying to reach specific code sooner), thus wasting too much energy on program paths that will not reach sinks.

In summary, Java is one of the most popular language suffering devastating ODD vulnerabilities \cite{In-Depth}, but there are few solutions to discover Java ODD vulnerabilities while existing solutions have limited efficiency and effectiveness. 
To address these challenges, we propose a novel hybrid solution \sysname to discover ODD vulnerabilities for Java applications.
In particular, \sysname performs a lightweight taint analysis, which makes a trade-off between precision and recall when handling Java runtime polymorphism, to identify possible candidate gadgets chains. Then, \sysname models the data constraints of such gadget chains as a tree and utilizes it to perform structure-aware fuzzing.
Finally, \sysname adopts a novel directed fuzzing solution driven by a step-forward mutation strategy and a hybrid feedback, to reach candidate vulnerabilities rapidly.

% \textcolor{blue}{organizes the multilevel class hierarchy of a target gadget chain as a tree structure and converts this tree into a syntactically valid injection object for fuzzing.} Furthermore, \sysname employs a step-forward mutation strategy, as well as both seed distance and gadget coverage as hybrid feedback to guide the fuzzer to mutate and schedule the injection objects towards the target sink rapidly, which further improves the efficiency of validation.

We implemented \sysname based on a popular Java fuzzing framework JQF \cite{JQF} and evaluated it on ysoserial \cite{ysoserial} which is a famous Java deserialization repository with 34 known gadget chains. As the evaluation results show, \sysname can identify 16 exploitable gadget chains without false positives, while two state-of-the-art solutions GadgetInspector and SerHybrid only respectively found three and two of them.
% Furthermore, the structure-aware input generation and feedback-driven fuzzing guidance also \textcolor{blue}{bring improvements in the validity and efficiency of gadget chain validation.} 
% Compared to state-of-the-art automated gadget chain discovery tools, \sysname hunts more exploitable gadget chains than GadgetInspector and SerHybrid, including 13 unique gadget chains that cannot be found by them. 
% Regarding the capability of detecting vulnerability, 
\sysname also identifies six previously unknown exploitable gadget chains in four popular Java applications. We have reported these vulnerabilities to the vendors and are working with them on fixing these vulnerabilities. In total, five of these six vulnerabilities have been assigned with new CVEs.

In summary, this paper makes the following contributions:
\begin{itemize}[leftmargin=1em]
\item We propose a novel solution to Java ODD vulnerability discovery, i.e., \sysname, which adopts a lightweight taint analysis to identify as many gadget chains as possible and a directed fuzzing solution to validate true positive chains.

\item We propose a step-forward and a structure-aware scheme to efficiently guide directed fuzzing towards sensitive sinks. 

\item We discovered and responsibly report six previously unknown exploitable gadget chains (i.e., ODD vulnerabilities).

\item We will open source our tool \sysname\footnote{https://github.com/ODDFuzz/ODDFuzz} to facilitate further research.

    % \item \textcolor{blue}{\textbf{A Gadget Chain Hunting Approach:} We present the design and implementation of \sysname, a novel gadget chain hunting approach designed to identify suspicious Java deserialization gadget chains and generate valid injection objects for validation.}
    
    % \item \textcolor{blue}{\textbf{A Directed Greybox Fuzzer}: We propose a new directed greybox fuzzer for gadget chains,} which combines step-forward mutation and hybrid feedback to efficiently guide the generated object to evolve towards the target sink.
    
    % \item \textbf{Discovered Vulnerabilities:} As part of our evaluation, we discovered six previously unknown exploitable gadget chains with five CVEs assigned in four popular Java applications. We responsibly disclosed the relevant details to the corresponding vendors.
    % \item \textbf{An Open-Source Tool:} We have open-sourced \sysname \footnote{https://github.com/CHASER-SP2023/CHASER}, in order to facilitate further research.
\end{itemize}

\section{Background}

% \textcolor{blue}{This section provides the background about open dynamic deserialization vulnerabilities in Java} (Section \ref{DV}) and discusses preliminaries about directed greybox fuzzing techniques (Section \ref{DGF}).

\subsection{Open Dynamic Deserialization}\label{DV}

\emph{Open Dynamic Deserialization} (ODD), also known as \emph{Object Injection Vulnerabilities} (OIVs) or insecure deserialization~\cite{owasp}, refers to a security-critical bug that allows an attacker to manipulate serialized objects to inject harmful data into the application code. This insecure deserialization behavior enables diverse attacks, including denial of service (DoS) attacks, or even remote code execution (RCE) \cite{Svoboda}. ODD occurs not only in Java, but also in other mainstream programming languages like JavaScript \cite{javascript}, PHP \cite{FUGIO,DBLP:conf/ccs/DahseKH14}, and .NET \cite{.Net}.

\noindent\textbf{Object Deserialization.}
Object serialization is a dynamic process of converting objects into a flatter format that can be sent and received as a sequential stream of bytes, for cross-platform data transmission and persistence storage. Object deserialization is the exact opposite of serialization, that is, restoring this byte stream to the original object. In other words, the object's properties are preserved along with their assigned values in the process of serialization and deserialization.

\begin{figure}[t]
\centering
\includegraphics[width=.95\linewidth]{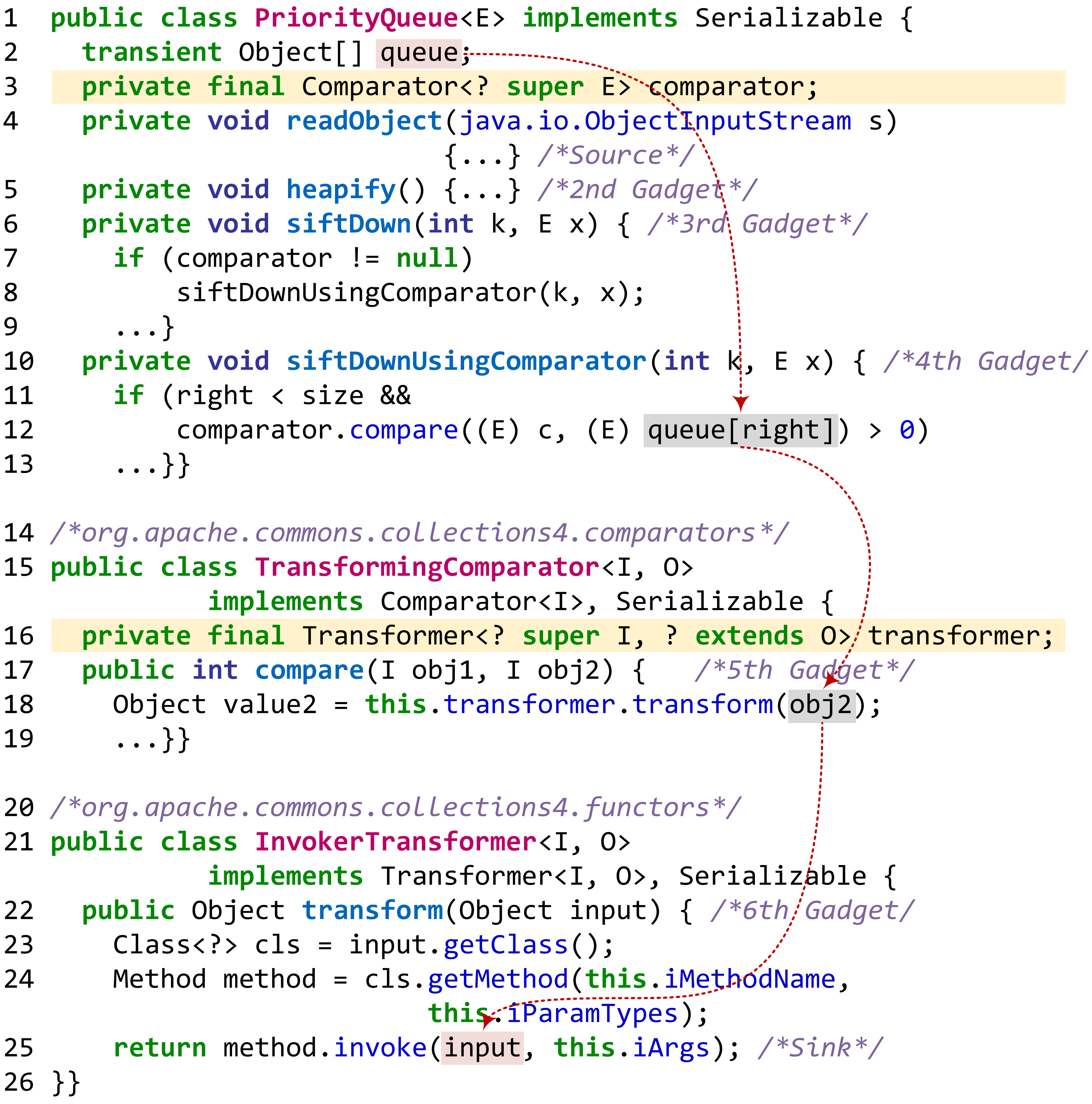}
\caption{\label{popcode}An exemplary Java ODD vulnerability.}
% \caption{\label{ExpObj}An example of constructing an injection object with POP to perform the deserialization attack.}
\vspace{-2mm}
\end{figure}

% Although object serialization and deserialization seems an innocuous mechanism, this 
Such a deserialization mechanism is \textit{open}, i.e., allowing arbitrary objects to be deserialized, and \textit{dynamic}, i.e., able to invoke polymorphic methods or reflection-based behaviors and explore diversified paths.
These two features can introduce serious ODD vulnerabilities \cite{DBLP:conf/ccs/HolzingerTBB16}. Typically, deserialized objects are assumed to be trustworthy after some checks. However, a large Java application may implement different libraries with their own dependencies. For developers, this creates a massive pool of classes and methods that are difficult to manage securely, because it is hard to predict which methods can be invoked by the malicious data due to the dynamic nature of Java. From the attacker’s perspective, deserializing data from any provenance provides an entry point to an object injection attack, if an attacker is able to chain code fragments of the application together (and execute them in order) and passes data to a security-sensitive call site. Such a code fragment chain is called a \emph{gadget chain}, and each code fragment of this chain is called a \emph{gadget}. Figure~\ref{popcode} shows a simplified code snippet of \code{CommonsCollections2}, a well-known gadget chain in \code{Apache Commons Collections4} (ACC) library \cite{ysoserial},  which enables remote code execution. 

\begin{figure}[t]
\centering
\begin{subfigure}{\linewidth}
\centering
\includegraphics[width=.8\linewidth]{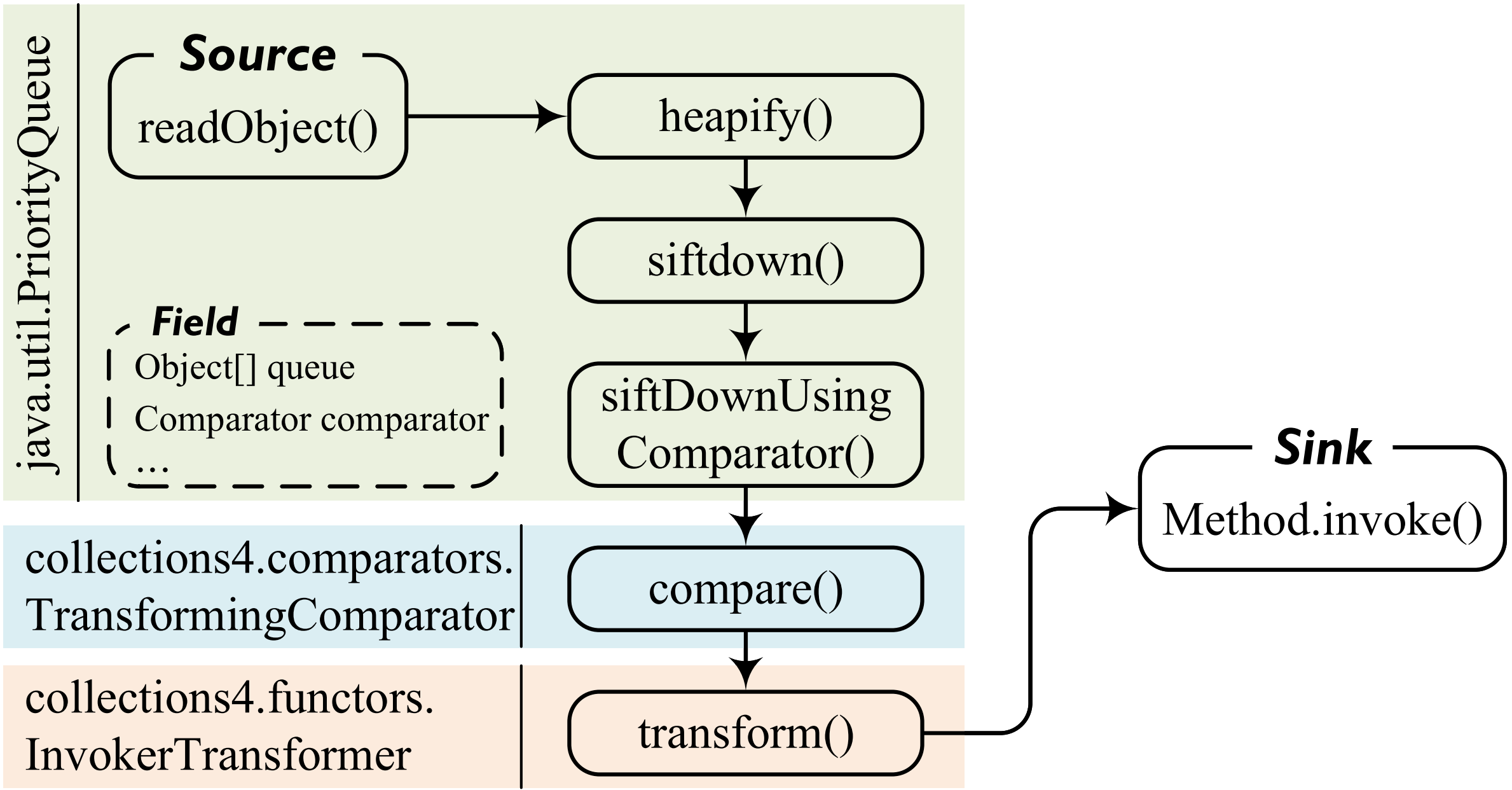}
\caption{\label{popchain}The stack trace of the gadget chain in Figure \ref{popcode}.}
\quad
\end{subfigure}
\begin{subfigure}{\linewidth}
\centering
  \includegraphics[width=0.8\linewidth]{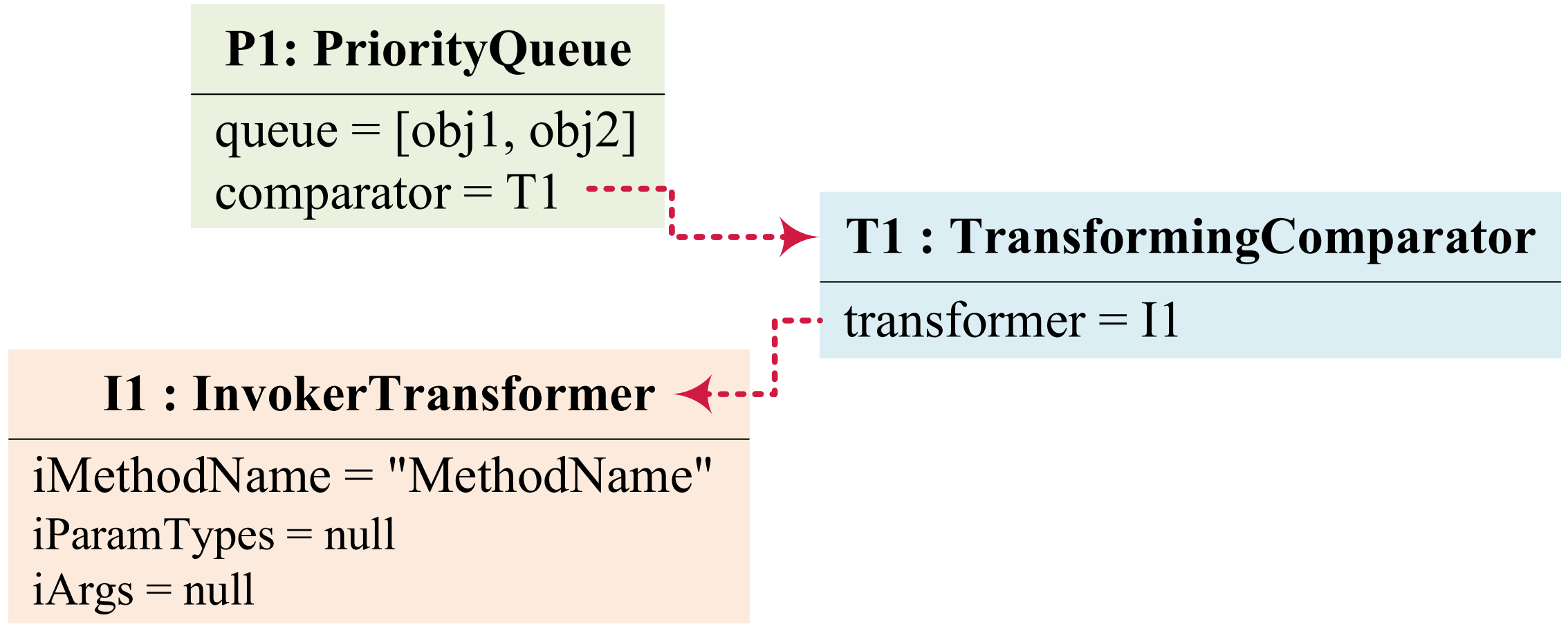}
\caption{\label{exploit}An injection object triggering \code{Method.invoke()}.}
\end{subfigure}
\caption{\label{POP}Constructing injection objects with POP.}
\end{figure}

\noindent\textbf{Property-Oriented Programming.}
% Unlike other types of vulnerabilities, an important factor of exploiting the object deserialization vulnerability is that the attackers could inject a crafted object, which joins reusable code fragments in memory (i.e., gadgets) together piece by piece (i.e., gadget chain) during deserialization, into a target application scope to perform malicious behaviors. 
To exploit ODD vulnerabilities, adversaries have to carefully set the properties of the injection object, to chain multilevel objects of specific classes and set certain fields to specific data values, in order to invoke specific polymorphic methods and pass data to security-sensitive call sites.
Such a technique used in constructing this injection object is called \emph{Property-Oriented Programming} (POP) \cite{POP}. POP allows an attacker to manipulate the data and control flow of the application, thereby exploiting attacker-controllable gadgets on the application's classpath for deserialization attacks.

% \begin{figure}[t]
%   \centering
%   \includegraphics[width=0.9\linewidth]{exploit.pdf}
% \caption{\label{exploit}An exploit object triggering \texttt{Method.invoke()}.}
% \label{POP}
% \end{figure}

Figure \ref{POP} depicts an example of how an attacker constructs a malicious injection object with POP to exploit the ODD vulnerability shown in Figure~\ref{popcode}, where Figure \ref{popchain} presents its corresponding stack trace. 
An attacker instantiates an injection object \code{PriorityQueue}, which contains a malicious payload within its field \code{queue} (line 2), for exploitation. At the bottom of Figure~\ref{popchain} are two exploitable gadgets, \code{compare()} (line 17) and \code{transform()} (line 22) in ACC, required to trigger the security-sensitive call site \code{Method.invoke()} (line 25). To enable the injection object to follow the execution flow that the gadget chain specifies, the attacker should dynamically set the property \code{comparator} (line 3) of \code{PriorityQueue} to an instantiated \code{TransformingComparator} object, and iteratively sets \code{TransformingComparator}'s property \code{transform} (line 16) to another instance \code{InvokerTransformer} to facilitate the payload \code{object} in \code{queue} reaching the sink, as shown in Figure~\ref{exploit}. When this crafted injection object \code{PriorityQueue} is deserialized via \code{readObject()}, 
% one of the most common entry point in object deserialization, 
the payload \code{object} in \code{queue} will flow into the security-sensitive call site \code{Method.invoke()}, thereby allowing remote code execution.

\subsection{Threat Model}
\begin{figure}[t]
  \centering
  \includegraphics[width=0.8\linewidth]{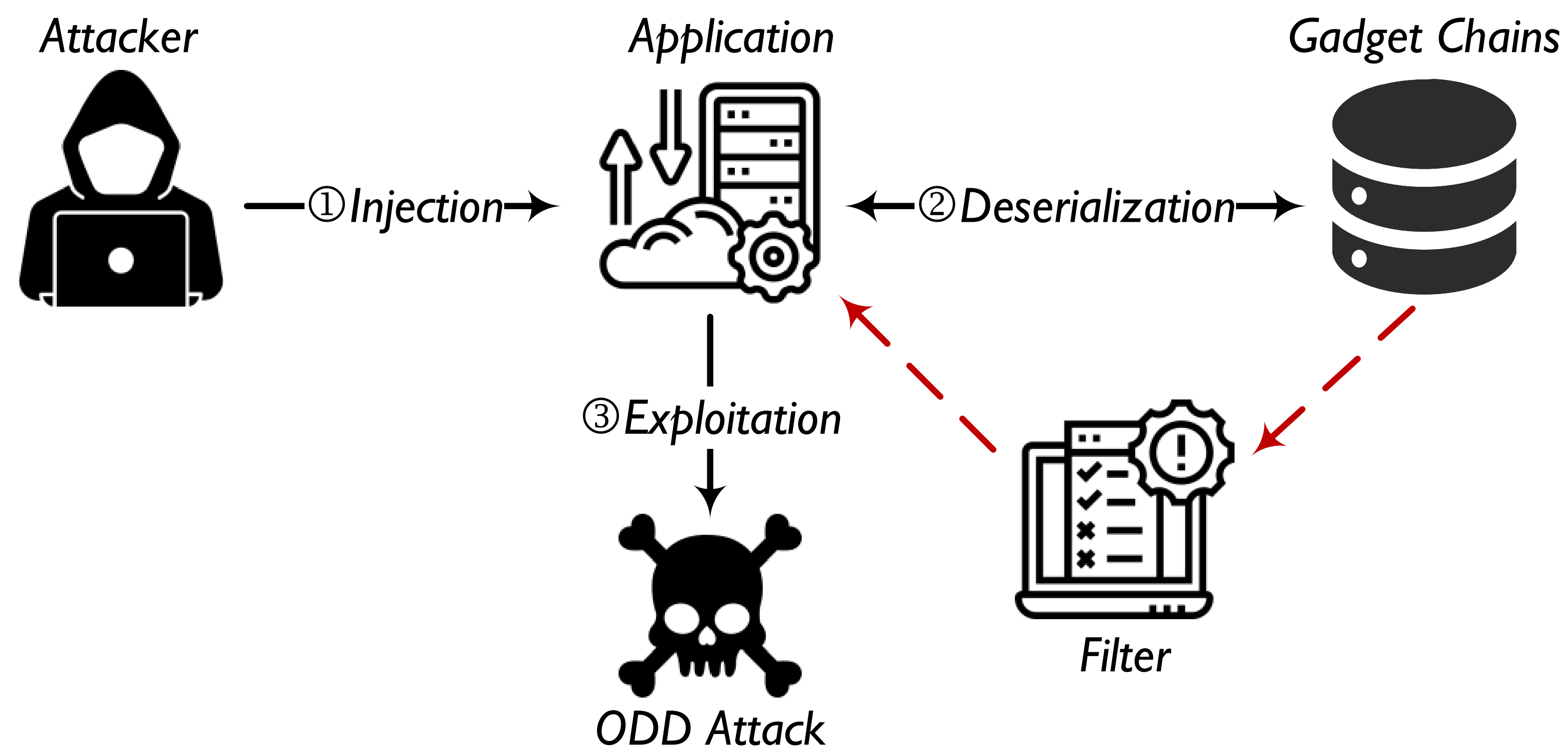}
  \caption{Threat model.}
  \label{ThreatModel}
\end{figure}

Figure \ref{ThreatModel} illustrates the threat model that we explored in this paper. Assume that there are attacker-controllable \emph{deserialization entry points} in the target Java application. If an attacker \ding{182} injects crafted objects into these untrusted entry points, the target application will \ding{183} deserialize these objects and automatically invoke attacker-specified gadget chains on the application's classpath. Then, \ding{184} malicious payloads carried by injection objects will flow into the security-sensitive call sites, enabling attackers to perform ODD attacks for exploitation.

This assumption is practical because insecure deserialization is common in the Java ecosystem. Take the prevalent commercial platform \code{Oracle WebLogic Server} (WLS) \cite{WebLogic} as an example. The T3 protocol WLS adopts to transport serialized data with other Java programs exposes a large attack surface, which provides untrusted deserialization entry points to attackers for sending payloads to the victim application. Nonetheless, containing deserialization entry points does not mean that the target application is vulnerable. As shown in Figure \ref{ThreatModel} (red dotted arrows), given that remediating an ODD vulnerability can be particularly difficult and costly, developers prefer adopting whitelists or blacklists to restrict the deserialization of untrusted objects \cite{Inspector,Carettoni,JEP290}. However, once a new gadget chain is discovered, existing defense solutions can be easily bypassed \cite{blackhat2019}. Thus, in our threat model, information about whether there are exploitable gadget chains on the application's classpath is more important.

\subsection{Directed Greybox Fuzzing}\label{DGF}
\begin{figure}[t]
  \centering
  \includegraphics[width=0.95\linewidth]{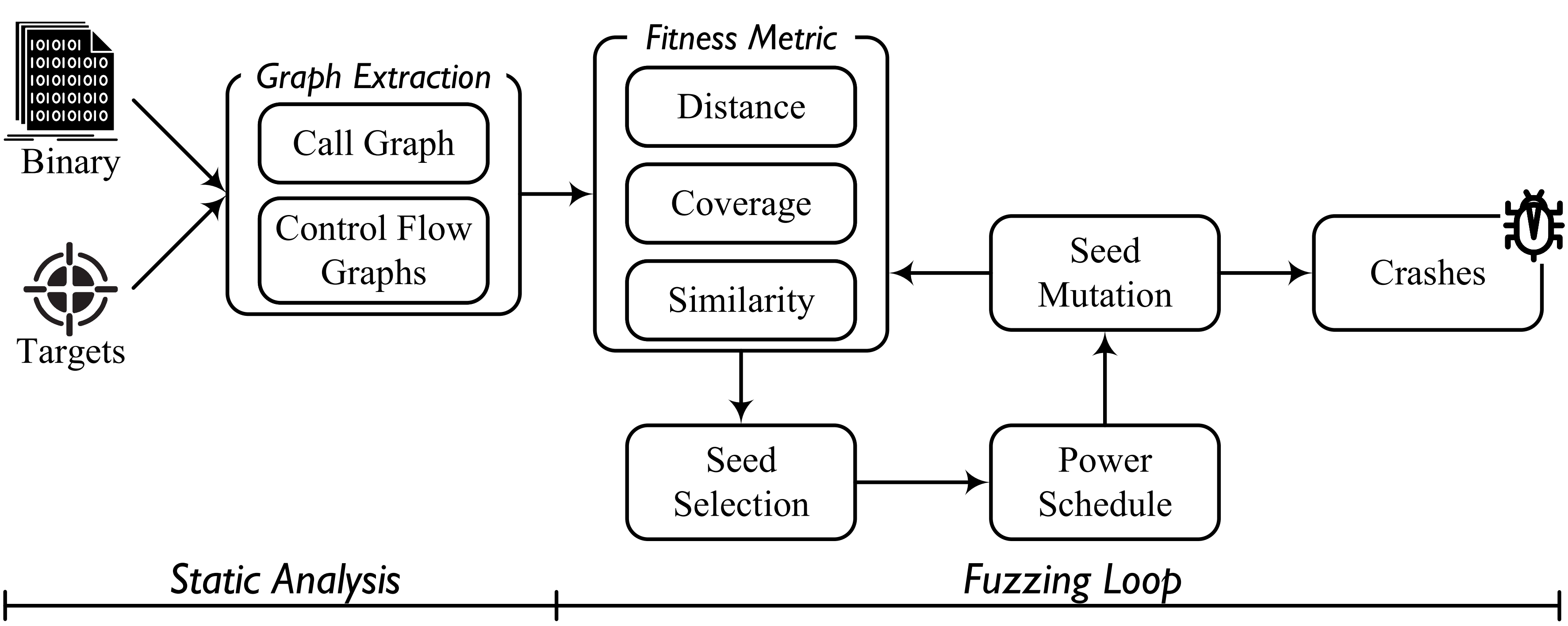}
\caption{Workflow of directed greybox fuzzing.}
\label{Workflow}
\end{figure}
Greybox Fuzzing (GF) has become an effective method to detect vulnerabilities \cite{fuzz}.  With different goals, it can usually be divided into two types: \emph{Coverage-guided Greybox Fuzzing} (CGF) \cite{CGF1,CGF2,CGF3} and \emph{Directed Greybox Fuzzing} (DGF) \cite{DGF,Hawkeye,BEACON}. CGF aims to explore previous undiscovered code snippets to achieve high code coverage, expecting to accidentally trigger potential vulnerabilities. However, in some scenarios, such as static report verification \cite{DBLP:conf/icse/Christakis0W16}, the vulnerable code is known and demands exploration. Hence, DGF are designed to guide the fuzzer to a specific location of the program to generate a \emph{Proof-of-Concept} (PoC) testcase.

\begin{table}[t]\scriptsize
 \caption{State-of-the-art automated gadget chain discovery tools. Intra-TA and Inter-TA respectively represent the intraprocedural and interprocedural taint analysis, while PTA denotes the points-to analysis.}
  \centering
  \renewcommand\arraystretch{1.2}
  \renewcommand\tabcolsep{2pt}
  \begin{tabular}{lm{1.4cm}<{\centering}m{1.4cm}<{\centering}m{1.5cm}<{\centering}m{1.7cm}<{\centering}}
    \toprule
    \textbf{Technique} & \textbf{Static Analysis} & \textbf{Seed Generation} & \textbf{Seed Mutation} & \textbf{Seed Prioritization} \\
    \midrule
    GadgetInspector \cite{Inspector}  &  Intra-TA &  - &  -  &  -\\ 
    SerHybrid \cite{Serhybrid}  & PTA  & Heap graph  & - &  -\\
    FUGIO \cite{FUGIO}   & Inter-TA & Property tree & Heuristics    & Feedback-driven \\
    \midrule
    \textbf{\sysname}  &  Intra-TA  & Property tree  & Step-forward & Target-directed \\
    \bottomrule
  \end{tabular}
  \label{fuzzers}
\end{table}

Figure \ref{Workflow} describes the workflow of the directed greybox fuzzing. Typically, DGF can be split into two phases: static analysis and fuzzing loop. At static analysis phase, the directed fuzzer extracts both the call graph and control flow graphs of the program to calculate the inter-procedural distance \cite{DGF} between the input binary and pre-defined targets. At the fuzzing loop phase, target distance is usually used as feedback information along with other fitness metrics like coverage \cite{IJON,UAFL} and similarity \cite{Hawkeye} to rapidly guide the fuzzer towards the target sites. Then, the directed fuzzer selects the seeds closer to the target sites in the seed pool according to feedback information and allocates proper \emph{energy} (i.e., power scheduling) for mutation. The energy of a seed determines how many new seeds can be generated. Then, the fuzzer adopts various mutation strategies to steer the seeds to evolve towards the desired target sites and executes the instrumented program. A new seed with smaller distance will be preserved for the next fuzzing loop.

\section{Motivation}\label{Motivation}

Despite the severe impact of insecure deserialization in practice, existing efforts on automatically discovering Java ODD vulnerabilities (especially exploitable gadget chains) are still unsatisfactory. For example, the state-of-the-art Java gadget chain discovery tool GadgetInspector \cite{Inspector} can only report few exploitable gadget chains in real-world applications. In fact, as we will show in the rest of this section, to achieve both a high recall (identifying more possible gadget chains) and precision (confirming more exploitable gadget chains), a Java ODD vulnerability discovery solution has to tackle three fundamental challenges.

\subsection{Challenge 1: Runtime Polymorphism}\label{challenge1}

The root cause of ODD vulnerabilities is that, the untrusted deserialized objects can \emph{reach} (in terms of control flow) and \emph{affect} (in terms of data flow) the security-sensitive call sites of target applications. Hence, existing works \cite{Inspector,Serhybrid} use static analysis to identify a combination of available gadgets in the code that can be exploited by attackers to customize insecure deserialization paths. However, due to the runtime polymorphism of Java language, virtual method invocations cannot be determined based on the declared types. As a result, it is difficult to precisely infer program paths that would be taken at runtime, resulting in a high false-negative rate.

A straightforward way is to perform \emph{Class Hierarchy Analysis} (CHA) \cite{CHA} to take a comprehensive view of both explicit and implicit method invocations. Unfortunately, blindly considering \emph{all} available gadgets on the application's classpath will inevitably lead to path explosion because the number of candidate gadgets increases exponentially as length increases. Hence, as shown in Table \ref{fuzzers}, GadgetInspector \cite{Inspector} respectively computes passthrough data flows from method arguments to 1) return values and 2) method invocations, and enumerates all available methods based on the class inheritance hierarchy and method overriding hierarchy to chain exploitable gadgets. However, given that the attacker-controllable property \cite{DBLP:journals/ese/ZhouBWSZLZC22} can propagate from a tainted argument to its subclass arguments not tracked, a set of exploitable gadgets will be missed by GadgetInspector since the Java runtime polymorphism is not considered in its intraprocedural taint analysis. FUGIO \cite{FUGIO} computes interprocedural data flows on its built depth-bounded call tree to prune infeasible gadget chains. However, when applied to Java ODD gadget chain discovery, this solution may not work because a typical Java application might integrate hundreds of libraries with their own dependencies. This creates a massive pool of classes and methods, making the call tree too deep and breadth to deal with.

\subsection{Challenge 2: Structured Input Construction}\label{challenge2}

To invoke a series of exploitable gadgets on the application's classpath, the structure of injection objects is often organized as a nested form with multilevel sub-objects. Still taking the gadget chain in Figure \ref{popcode} as an example. To trigger the gadget \code{compare()} (line 17), the fuzzer should instantiate the class \code{TransformingComparator} to which the overridden method \code{compare()} belongs and assign this instance to the field \code{comparator} (line 3) of the injection object \code{PriorityQueue} through POP. This brings the challenge to constructing a both syntactically (i.e., the generated injection object can be (de)serialized) and semantically (the generated injection object satisfies certain control- and data-flow constraints that enable the gadget chain) valid fuzzed input, as it requires 1) shaping the injection object’s multilevel hierarchy to enable the execution of the reported gadget chain, and 2) assigning proper property values to trigger the security-sensitive call site. Without prior knowledge about such a complex nested form of object structures, traditional fuzzing techniques cannot thoroughly fuzz the entire gadget chain as they hardly figure out complex structures behind each injected object.

\begin{figure}[t]
  \centering
  \includegraphics[width=0.7\linewidth]{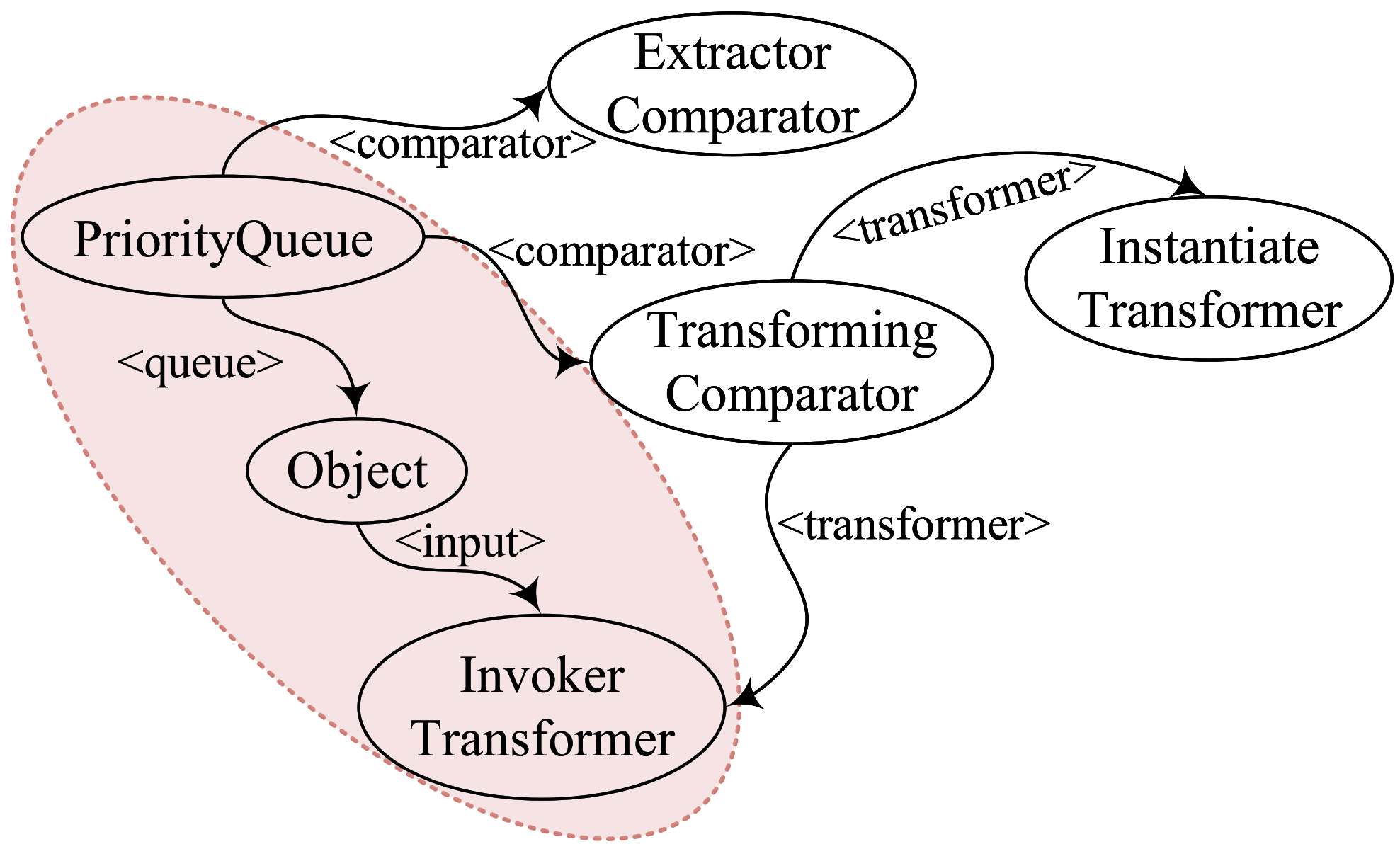}
\caption{\label{HeapGraph}The heap graph for the gadget chain in Figure \ref{POP}.}
\end{figure}

An effective solution to handle such nested object structures is generation-based fuzzing techniques. As shown in Figure \ref{HeapGraph}, SerHybrid \cite{Serhybrid} performs points-to analysis to produce a heap access path (pink-shaded), which satisfies the data-flow constraints of reaching the security-sensitive call site, from the heap graph and uses fuzzing to assign random values to the field properties not appear in the heap path according to their types to generate valid injection objects for execution. However, as the number of available gadgets increases, a fuzzer unaware to the multilevel class hierarchy of the injection object may hard to assign proper values to those control-data constraints-related properties. For example, it is unlikely to select \code{TransformingComparator} from a large number of implementations of the interface \code{Comparator}, resulting in runtime exceptions. FUGIO \cite{FUGIO} builds a property tree based on the candidate gadget chain to satisfy the control-flow constraints of sink-reachable injection objects, and mutates each property with some heuristic rules to construct actual injection objects. However, the random combination of arbitrary sub-objects (property values) generated by fuzzing may be semantically (hard to trigger target gadgets) invalid, blocking the injection objects from reaching gadgets closer to target sinks.

\subsection{Challenge 3: Target-Directed Fuzzing}\label{challenge3}

\iffalse
\begin{figure}[t]
  \centering
  \includegraphics[width=0.9\linewidth]{ExecutionTrace.png}
\caption{\label{ExecutionTrace}Execution traces of different seeds.}
\end{figure}
\fi

Since the gadget chain is composed of a series of attacker-controllable methods which are automatically executed during object deserialization, the conventional code coverage is not suitable to guide the fuzzer because a generated injection object which triggers more code snippets may not be able to reach the security-sensitive call site of the target chain. For example, an injection object whose property \code{comparator} is \code{null} (line 7 in Figure \ref{popchain}) will be preserved by the coverage-guided fuzzer (e.g., FUGIO) as an interesting seed for the next fuzzing loop since it triggers new code snippet (line 13) in the gadget \code{siftDown()} (line 6). As a result, the fuzzer will waste most of its time budget on exploring unreachable paths. 

Instead of focusing on maximizing the code coverage, DGF prioritizes the seeds whose execution traces are close to the target sites to gain directedness. State-of-the-art directed fuzzers leverage the arithmetic mean of the distances of all the basic blocks on a seed's execution trace to select and schedule the seeds to reach target sites rapidly. However, such a seed distance can be biased and may not entirely correspond to the expected execution path of a gadget chain being validated since not every block drives the seed object to execute towards the target sink expected in an identified chain. Moreover, the execution traces of different seed objects may vary greatly and can only be known at runtime since modifications to a property of the seed object may activate the execution of multiple gadgets. Hence, target-directed fuzzing feedback that can effectively evaluate the quality of generated injection objects is desired.

\section{ODDFuzz Design}
To tackle the aforementioned challenges of gadget chain discovery discussed in Section \ref{Motivation}, we design \sysname to support structure-aware directed greybox fuzzing. In this section, we outline the overall design and workflow of \sysname, and explain its key components.

\subsection{Overview}

\begin{figure}[t]
  \centering
  \includegraphics[width=\linewidth]{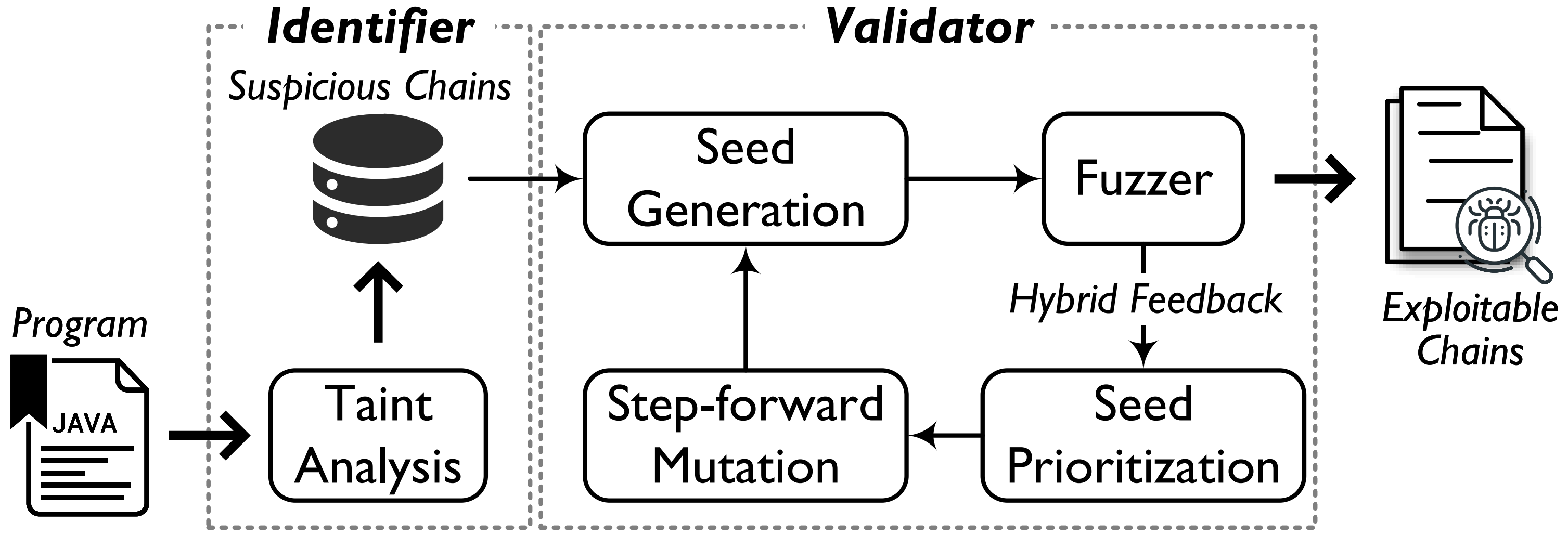}
\caption{Overview of \sysname.}
\label{Fuzzing Loop}
\end{figure}

Figure \ref{Fuzzing Loop} depicts an overview of \sysname. The workflow of \sysname contains two main modules: \ding{182} In the \emph{identifier} module, \sysname takes a compiled file (e.g., Jar, War, or Class files) of the \emph{program under testing} (PUT) as input and conducts a lightweight taint analysis to automatically enumerate all potential gadget chains. \ding{183} In the \emph{validator} module, \sysname generates a structure-aware seed based on the identified gadget chains to construct syntactically valid injection objects for fuzzing. During the fuzzing loop, \sysname combines step-forward mutation strategy and hybrid feedback (seed distance and gadget coverage) to guide the fuzzer to mutate injection objects towards the desired sinks. When a generated injection object reaches the security-sensitive call site, the given gadget chain will be reported as an exploitable gadget chain.

\subsection{Taint Analysis}

An important condition for constructing an exploitable gadget chain is that whether the attacker-controllable tainted object can propagate from an entry point (i.e., source) to the security-sensitive call site (i.e., sink) method. In other words, if an exploitable gadget chain exists, there must be a call path from the entry point to the security-sensitive call site. A straightforward way is to construct Call Graph (CG) \cite{CG,DBLP:conf/icse/CaoSBWLT22} to search for reachable paths \cite{DBLP:conf/sigsoft/Cheng0BW22}. However, due to the Java runtime polymorphism, virtual method invocations cannot be determined based on the declared types. To solve this problem, we perform a lightweight summary-based taint analysis \cite{FlowDroid,StubDroid,DBLP:conf/popl/TangWZXZM15} to identify suspicious gadget chains.

\noindent\textbf{Method Summary Computation.}
\sysname first computes static summaries for all methods on the classpath of the PUT that are later used for constructing gadget chains. 

Specifically, for each method, \sysname first extracts all its arguments and \code{this} as method summaries. Then, to track the information propagation between variables of each method, we focus on four basic statements, including 1) Assign, 2) Load, 3) Store and 4) Call. These statements are widely used for data flow computation in taint analysis. The variable that data-dependent on an argument of the method will also be included in the method's summaries. These method summaries will be used to identify exploitable gadgets whose actual arguments can be controlled by attackers to propagate tainted values by altering the property values of an injection object.

\noindent\textbf{Gadget Chain Identification.}
Since the gadget chain is a sequence of method invocations that reflects a stack trace from a magic method to a security-sensitive call site, \sysname should specify a list of exploitable magic methods and security-sensitive call sites, and identify suspicious gadget chains based on previous computed method summaries. In this paper, we specified a total of 16 magic methods and 30 security-sensitive call sites (as shown in \refappendix{Source&Sinks}).

Then, given that the Breadth-first-search (BFS) adopted by GadgetInspector will skip visited methods (i.e., gadgets which have been traversed on certain infeasible paths will not be considered for gadget chain construction again even if they are exploitable) and thus results in false negatives, once a known magic method is found on the classpath of the PUT, \sysname performs a Depth-first-search (DFS) starting from this source gadget based on the method summaries to chain exploitable gadgets. To avoid infinite loops (e.g., recursive calls), we set a threshold for the maximum length of candidate gadget chain. Furthermore, to handle the runtime polymorphism of Java language, we perform Class Hierarchy Analysis (CHA) on the call statement only when the caller is tainted, avoiding the path explosion issue caused by blindly considering \emph{all} available gadgets on the application’s classpath. In particular, for a call statement $r = x.k(a, \cdots)$, if the caller variable $x$ is tainted (e.g., \code{comparator.compare()} at line 12 in Figure \ref{popcode}), all overriding methods of method $k$ will be listed as candidates. Otherwise, \sysname works like a normal CG-based taint analyzer. This iterative analysis procedure will not stop until a security-sensitive sink method is invoked or the maximum length of the enumerated gadget chain exceeds a threshold.

After all paths (i.e., gadget chains) from magic methods to security-sensitive call sites are analyzed, \sysname runs the validator module for validation. With the help of our lightweight taint analysis, the effectiveness (identifying as many gadget chains as possible) and scalability (analyzing large applications with acceptable time overhead) of \sysname in gadget chain identification can be well balanced.

\subsection{Structure-Aware Directed Greybox Fuzzing}
Given a target Java application and a candidate gadget chain, \sysname conducts structure-aware directed greybox fuzzing to generate actual injection objects for validation. The main fuzzing loop is as presented in Algorithm \ref{algorithm} in \refappendix{AlgorithmDetails}, which is composed of the following three main components.

\noindent\textbf{Structured Seed Generation.}
As described in Section \ref{challenge2}, constructing a syntactically valid injection object requires 1) devising its nested object hierarchy that reflects the execution flow of a given gadget chain, and 2) assigning suitable property values to corresponding multilevel sub-objects to facilitate the injection object reaching the sensitive sink. However, heavy use of nested structures makes gadget chain fuzzing ineffective, as it requires well-designed property layout of complex object structures, which is unfriendly to traditional fuzzing solutions.

\begin{figure}[t]
  \centering
  \includegraphics[width=0.9\linewidth]{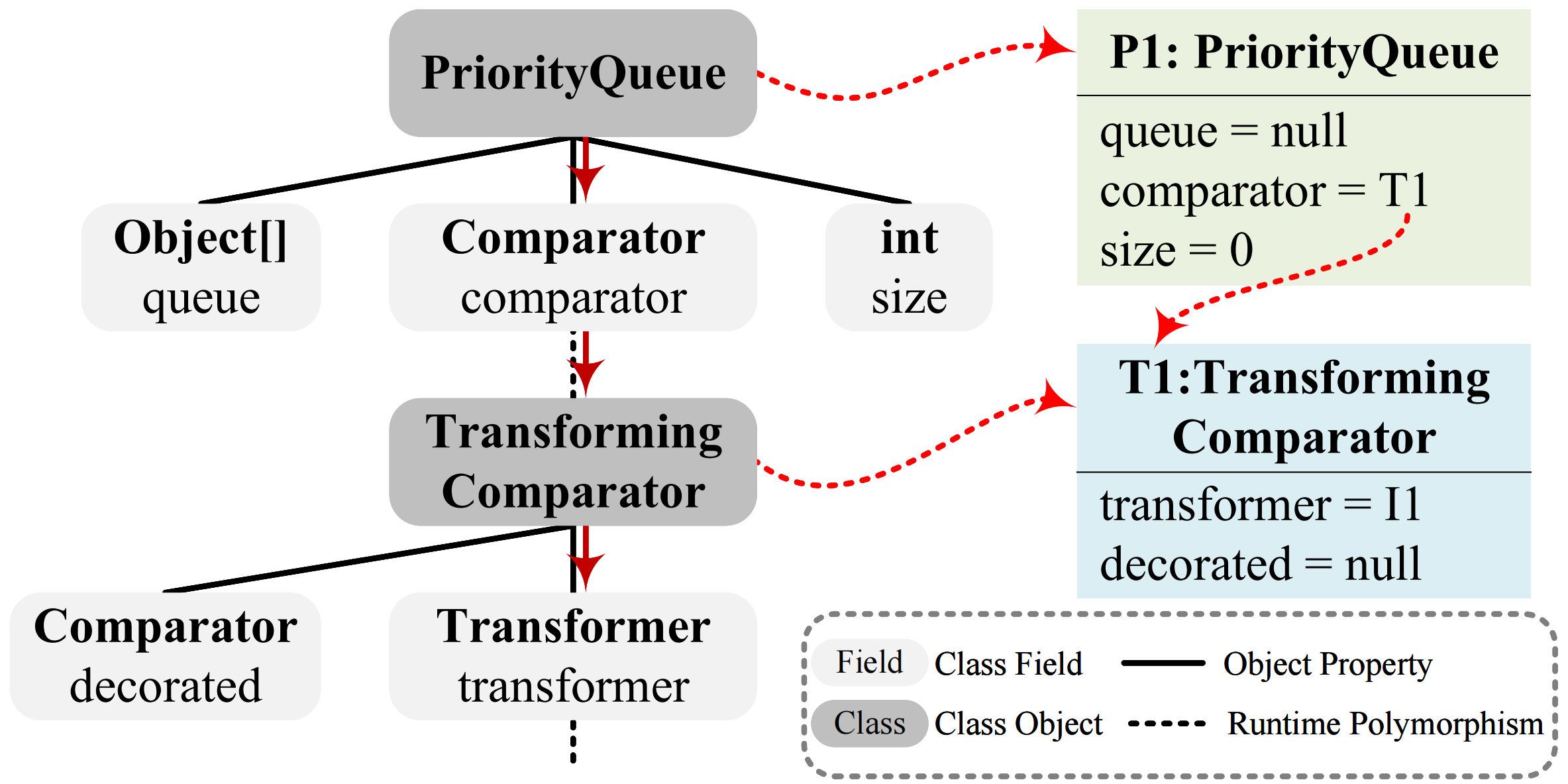}
\caption{A merged property tree for injection object generation.}
\label{PropertyTree}
\vspace{-2mm}
\end{figure}

To this end, we design a structure-aware seed generation approach to handle the complex nested forms by adopting a hierarchical data structure called \emph{property tree} \cite{FUGIO}, in which the root node represents a class object that holds one or more gadgets, and leaf nodes are a series of class fields which contain the property type and name. As shown in Figure \ref{PropertyTree}, to generate an injection object shown in Figure \ref{POP}, we instantiate each class involved in the gadget chain and leverage reflection to dynamically collect available properties of each class to construct a property tree. Specifically, if the property type of a field node in a property tree is an object represented (or inherited) by another property tree of which the class holds the next gadget in the target chain, we merge the two property trees by connecting this field node to its corresponding class object node  (i.e., the root node of another property tree). It is noteworthy that two property trees are also merged when certain field node's type in a property tree is the interface implemented by the root node (a class object) of another property tree. For example, the property type of the field node \code{comparator} in the property tree of class \code{PriorityQueue} is the interface \code{Comparator}. Hence, the two property trees of class \code{PriorityQueue} and class \code{TransformingComparator} are merged by connecting the field node \code{Comparator comparator} and the root node \code{TransformingComparator}. We iteratively integrate the property tree based on the invocation order of the gadget chain until there are no more isolated but related sub-trees.

When a suspicious gadget chain is identified by \sysname, it will be fed into the input generator to construct a corresponding property tree. The multilevel class hierarchy of a target gadget chain can be well modeled with the property tree. Then, the fuzzer starts traversing the backbone of this tree to convert it into an initial injection object for fuzzing (the right side of Figure \ref{PropertyTree}). Other property nodes  without successors (e.g., \code{Object[] queue}) will be set to \code{null} for mutation.

\noindent\textbf{Seed Prioritization via Hybrid Feedback.}
Using the above-mentioned structure-aware seed generation that handles the complex nested forms, we can successfully construct syntactically valid injection objects to enable the gadget chain fuzzing process. However, as described in Section \ref{challenge3}, the execution trace of an injection object is dynamically determined, which means that randomly generating and mutating an injection object leads to the sink-unawareness since the property layout of this nested injection object varies greatly in different fuzzing iterations. Such an indeterministic fuzzing campaign without clear feedback guidance degenerates the fuzzer into a semantics-blind dumb fuzzer. As a result, the fuzzer would be confused about which direction to evolve and wastes time on exploring unreachable paths, resulting in low efficiency.

In order to efficiently select and schedule the seeds to reach the security-sensitive call site of a given gadget chain, we propose a hybrid feedback-driven seed prioritization way. Fundamentally, we aim to prioritize and assign more energy to seeds closer to the target security-sensitive call site for mutation. To this end, \sysname takes two types of feedback metrics into account: \emph{seed distance} and \emph{gadget coverage}.

\subsubsection{Seed Distance}
Computing seed distance to prioritize and schedule seeds to reach target sinks as rapid as possible is a core component of DGF. Following the idea of AFLGo \cite{DGF} and Hawkeye \cite{Hawkeye}, the distance between a seed $s$ and the target basic block $T_b$ to which the security-sensitive call site belongs is calculated as:

\begin{equation}
    d(s,T_b) = \frac{\sum_{m\in\xi(s)}d_b(m,T_b)}{|\xi(s)|}
\end{equation}
where $d_b(m,T_b)$ is the distance between a basic block $m$ in the execution trace of seed $s$ and the target basic block $T_b$. It is noteworthy that instead of enumerating all the basic blocks on the execution path of a seed $s$, we collect the executed basic blocks $\xi(s)$ within the gadgets of the target chain to compute the seed distance, avoiding the fuzzer exploring irrelevant but closer paths.

\subsubsection{Gadget Coverage}
Furthermore, we also adopt gadget coverage (i.e., branch coverage of gadgets in a target chain) as another metric to prioritize seeds which cover more program paths. In the initial fuzzing stage, the gadget coverage aims at guiding the fuzzer to select and prioritize diverse seeds, avoiding getting stuck in local optimum caused by favoring certain seeds with specific execution paths. While in the power scheduling stage, the gadget coverage attempts to give seeds with the same distance but covering more branches higher chances for mutation.

Formally, \sysname sorts all the generated seeds in ascending order according to their distance and maintains a two-level priority queue. The first seed (or seeds with same distance but different coverage) will be put into the favored queue with higher priority, and the rest of the seeds are put into the less favored queue. Thus, \sysname has a greater chance to select the next seed from the favored queue for mutation. As for power scheduling, \sysname uses Equation (\ref{power}) to consider both seed distance and gadget coverage to assign a proper energy to the selected seed input.

\begin{equation}\label{power}
    p(s,T_b) = \psi(s) \cdot (1 - \widetilde{d}(s,T_b))
\end{equation}
where $\psi(s)$ denotes the proportion of the branches of gadgets covered by a seed $s$ (i.e., gadget coverage) to the total branches of all gadgets in a given chain, and $\widetilde{d}(s,T_b) = \frac{d(s,T_b) - minD}{maxD - minD}$ is a normalized seed distance where $minD$ (or $maxD$) is the smallest (or largest) seed distance ever met. It is obvious that $p(s,T_b) \in [0,1]$ since both the multipliers are in $[0,1]$.

With Equation (\ref{power}), the fuzzer can determine the number of mutation chances to be applied on the current seed and evaluate whether the mutated seeds should be favored during the seed prioritization, striking a balance between exploring diverse execution paths and prioritizing a seed that is more likely to reach the desired security-sensitive call site.

\noindent\textbf{Step-Forward Seed Mutation.}
Previous fuzzing techniques work by randomly mutating binary files via operations like bit flips to produce new inputs. However, such bit-level mutations may lead to invalid syntax when applied to structured inputs. To address this issue, we leverage JQF \cite{JQF}, a parametric fuzzing framework which maps the structured inputs to a sequence of untyped bits (i.e., parameters), to mutate the generated seeds at the bit-level. These bit-level mutations on the parameters correspond to property-level mutations on structured injection objects. Then, \sysname applies a step-forward seed mutation strategy to efficiently guide the seeds towards the desired security-sensitive call site of a target gadget chain.

Specifically, the fuzzer first traverses the property tree of an injection object to be mutated and checks each property's type. For \emph{primitive} data types (e.g., \code{boolean}, \code{int}), the fuzzer uses multiple pseudo-random methods proposed by JQF to convert untyped bit parameters into random typed values. For the \emph{reference} data types, the fuzzer tailors targeted templates for specific types. When the property type is \code{class}, the fuzzer will randomly select a class from the candidate classes (i.e., sub-classes) of this property via the method \code{random.choose()}. For an \code{array} property, the fuzzer uses the method \code{random.nextInt()} to randomly set up the array size and assigns random values based on the type of elements (i.e., instances that inherit the class type of the array) to the array. For example, the parameter sequence for an injection object generated from the property tree in Figure \ref{PropertyTree} is:

\vspace{-0.2cm}
\begin{figure}[h]
  \centering
  \includegraphics[width=0.7\linewidth]{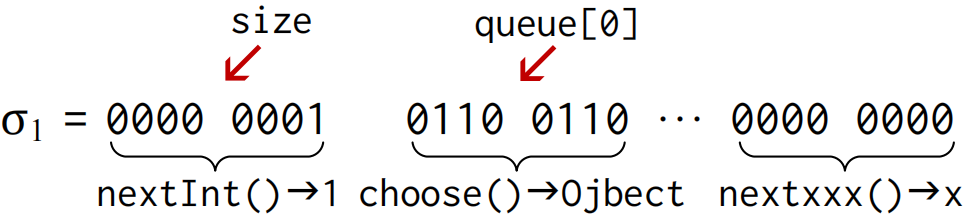}
\end{figure}
\vspace{-0.2cm}

In order to mutate the value of property \code{size}, which is a variable with \code{int} type in class \code{PriorityQueue}, the fuzzer invokes the method \code{random.nextInt()} to generate a random integer \code{1}. To generate an \code{Object} array \code{queue}, the fuzzer invokes the method \code{random.choose()} to assign it an instance \code{Object} from the pre-defined dictionary, which is composed of some specific property values (e.g., class object, string object) involved in \emph{all} classes or methods in the candidate gadget chain. These pre-defined values have a higher probability to satisfy certain hard dependencies during fuzzing.

Furthermore, to guide the seeds towards a desired sink method, \sysname mutates the nested sub-objects of the interesting injection object at the bit-level one by one. To this end, we insert additional \code{identifier} bytes with the method \code{random.nextBool()} into the parametric sequence of an injection object. When the fuzzer meets a class object node while traversing the property tree, the fuzzer adds a byte as an identifier to mark whether to mutate the property values of this nested sub-object. We leverage the gadget coverage collected by the fuzzer to identify the class where the last branch covered by the injected object is located. Once an injection object is stuck in certain gadgets, the fuzzer will set the corresponding identifier bytes to \emph{true} and assign random values to parameters, which correspond to structural mutations on the properties of the class to which the stuck gadget belongs, to produce new inputs.

To illustrate our step-forward mutation, considering the following parameter sequence $\sigma_2$:

\vspace{-0.2cm}
\begin{figure}[h]
  \centering
  \includegraphics[width=0.7\linewidth]{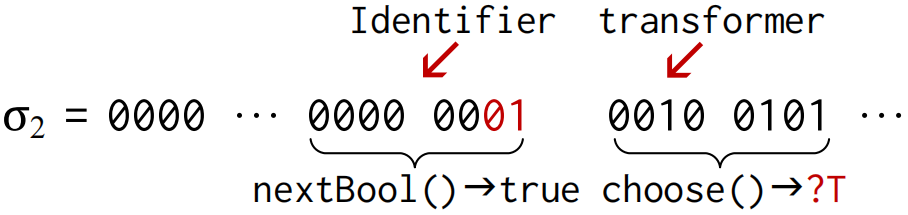}
\end{figure}
\vspace{-0.2cm}

Suppose that there is an injection object that stops in the gadget \code{TransformingComparator.compare()}, the fuzzer will flip its \code{Identifier} to \emph{true} and mutates the parameter sequence (e.g., assigning an instance \code{?T}\footnote{Due to space constraints, we use \code{?T} to represent an instance of \emph{any} class that implements \code{Transformer}.} to the property \code{transformer}) corresponding to class \code{TransformingComparator}. Based on this step-forward mutation strategy, the fuzzer can effectively generate semantics-aware inputs which are more likely to reach the target sink.

Finally, when the mutated seed reaches the security-sensitive call site, the fuzzer will report that the given gadget chain is exploitable with the generated injection object.

\section{Implementation}
We implemented \sysname  based on a popular Java fuzzing platform JQF \cite{JQF}. We customized its components to make it suitable for gadget chain fuzzing while piggybacking on the underlying functionalities of JQF, such as runtime instrumentation.

\noindent\textbf{Taint Analysis.}
\sysname uses Soot \cite{Soot} to parse and convert the Java bytecode to the intermediate language Jimple \cite{jimple}. Based on the basic class information (e.g., class modifier, field, method and instructions) from Jimple, we implemented the method summary-based taint analysis.

\noindent\textbf{Structured Fuzzing.}
Instead of manually writing declarative specifications of the input format such as context-free grammars or protocol buffers, \sysname modifies the \code{junit-quickcheck} generators \cite{junit-quickcheck} built in JQF to randomly generate and mutate structured injection objects based on the candidate gadget chains. To enable and facilitate the structure-aware seed generation, we employ the class \code{sun.msic.Unsafe} \cite{Unsafe} provided by JRE, allowing users to create an instance of a class without invoking its constructor code, initialization code, various JVM security checks and all other low level things.

\noindent\textbf{Runtime Instrumentation.}
We use the ASM toolkit \cite{ASM} to instrument Java bytecode on-the-fly via a javaagent as classes are loaded by the JVM. When the PUT starts, the \sysname instrumentor injects a static method invocation that is executed after each call or jump instruction to keep track of the execution trace of an injection object. Note that the instrumentation is limited to gadget chain-related bytecode instead of the whole program for efficiency concerns.

\noindent\textbf{Feedback Collection.}
For coverage information, we make minimal modifications to JQF to collect branch coverage through instrumenting each basic block based on jump instructions. For distance information, \sysname generates the corresponding intraprocedural control flow graphs (CFGs) of the gadget chain at the bytecode-level based on ASM. The root node (i.e., gadget) of a CFG is identified by the method signature while other CFG nodes are identified by the jump instructions of the corresponding basic blocks. When a gadget chain is fed to the fuzzer for validation, the \sysname distance calculator computes the inter-procedural distance towards the dangerous sink for each basic block based on the invocation order of the gadget chain and generated CFGs. The distance calculator is implemented with JGraphT library \cite{JGraphT}.

\section{Evaluation}\label{Evaluation}
In this section, we evaluate \sysname from different perspectives. First, we measure the effectiveness of \sysname for gadget chain identification, and demonstrate how the structure-aware seed generation and semantic-aware fuzzing guidance of \sysname contribute to triggering exploitable gadgets (Section \ref{Effectiveness}). Then, we compare its performance with state-of-the-art automated gadget chain identification tools, including an open-source tool and a previous study (Section \ref{ComparisonTools}). Finally, we show that \sysname can discover previously unknown vulnerabilities in popular Java applications (Section \ref{VulnerabilityHunting}).

\noindent\textbf{Experiment Environment.}
All experiments were conducted on a Linux workstation with an Intel(R) Core(TM) i9-12900k @3.90GHz and 256 GB of RAM, running Ubuntu 18.04.4 LTS with JDK 1.8.0\_152. 

\noindent\textbf{Benchmark.}
We performed the evaluation on various gadget chains based on  the ysoserial repository \cite{ysoserial}, a collection of 34 known gadget chains discovered in 22 common Java libraries that can be exploited to perform unsafe object deserialization.

\noindent\textbf{Exploitability Evaluation.}
To evaluate whether the gadget chains reported by \sysname and baselines were truly exploitable, we employed two professional security analysts to manually inspect each reported gadget chain. For 34 known gadget chains in the benchmark (Section \ref{Effectiveness} and \ref{ComparisonTools}), security analysts compared the gadget chains reported by each approach with the gadget chain attached to the payloads (ground truth) in ysoserial (e.g., as explicitly marked in the annotation of \code{CommonsCollections1} \cite{CommonsCollections1}). Given that the reported gadget chain may not be completely consistent with the ground truth (e.g., the gadget chain \code{CommonsCollections1} reported by GadgetInspector and ysoserial), the reported gadget chain would be confirmed as known if its core gadgets\footnote{In GadgetInspector, these application-specific continuous gadgets are also called \emph{the building block of the full gadget chain}.} involved in the vulnerable application/library were the same to those in ysoserial. For the remaining gadget chains (discovered in ysoserial (Section \ref{ComparisonTools}) and real-world applications (Section \ref{VulnerabilityHunting})), two security analysts manually inspect these reported gadget chains. Once a gadget chain was suspected to be exploitable, they would construct actual exploits for confirmation.

\subsection{Effectiveness}\label{Effectiveness}
To evaluate the effectiveness of \sysname, we repeated each experiment 10 times and reported their average statistical performance \cite{DBLP:conf/ccs/KleesRCW018}. We empirically set the threshold for each gadget chain to 15 gadgets. For each statically identified gadget chain, we limit the fuzzing campaign of \sysname to 120 seconds. We performed additional sensitivity analysis on these two hyperparameters for evaluation in \refappendix{Sensitivity}.

\subsubsection{Overall Performance}
\begin{table*}[t]\scriptsize
 \caption{Evaluation results of \sysname on known gadget chains from ysoserial. }
  \centering
  \renewcommand\tabcolsep{3.4pt}
  \begin{tabular}{lm{1cm}<{\centering}|m{1cm}<{\centering}m{1cm}<{\centering}m{1.2cm}<{\centering}m{1.1cm}<{\centering}m{1.1cm}<{\centering}m{1.1cm}<{\centering}m{1.4cm}<{\centering}m{1.4cm}<{\centering}m{1.1cm}<{\centering}m{1.2cm}<{\centering}}
    \toprule
    \textbf{Application}     & \textbf{Version} & \textbf{LoC} & \textbf{Classes} & \textbf{Methods}& \textbf{Covered Sources} & \textbf{Covered Sinks}   & \textbf{Known Chains}   & \textbf{Identified Chains}      & \textbf{Confirmed Chains}  & \textbf{Analysis Time} & \textbf{Fuzzing Time} \\
    \midrule
    JDK                & 1.7 & 4.4M & 38.5K & 324.6K & 7 & 4 & 4 & 9 (1) & 1 & 1m51s & 16m32s \\ 
    AspectJWeaver      & 1.9.2 & 692.4K & 7.1K & 19.8K & 4 & 2 & 1 & 9 (1) & 0 & 1m56s & 18m \\
    BeanShell          & 2.0b5        & 44.8K & 1.1K & 17K & 3 & 1 & 1 & 8 (0) & 0 & 1m53s & 16m \\
    C3P0               & 0.9.5.2      & 30.3K & 644 & 10.1K & 6 & 3 & 1 & 13 (1) & 1 & 1m50s & 25m53s \\
    Click              & 2.3.0 & 10.8K & 73 & 8.5K & 4 & 1 & 1 & 8 (1) & 1 & 1m48s & 15m26s  \\
    Clojure            & 1.8.0        & 58.4K & 3.8K & 25.7K & 5 & 4 & 1 & 184 (1) & 1 & 3m30s &  6h7m34s\\
    CommonsBeanutils   & 1.9.2        & 71.4K & 504 & 7.8K & 3 & 1 & 1 & 8 (1) & 1 & 1m52s & 14m25s \\
    CommonsCollections & 3.1          & 101K & 798 & 9.7K & 7 & 4 & 5 & 97 (5) & 3 & 1m58s & 3h10m53s \\
    CommonsCollections4& 4.0          & 101K & 630 & 7.4K & 5 & 2 & 2 & 112 (2) & 2 & 1m55s & 3h41m9s\\
    FileUpload         & 1.3.1        & 10.5K & 56 & 3.1K & 3 & 1& 1 & 8 (0) & 0 & 1m55s & 16m \\
    Groovy             & 2.3.9        & 252.4K & 4.2K & 45.6K & 4 & 1 & 1 & 13 (0) & 0 & 2m8s & 26m \\
    Hibernate          & 4.3.11        & 855.7K & 7.4K & 42.7K & 3 & 1 & 2 & 8 (2) & 2 & 2m8s & 14m7s \\
    JBossInterceptors  & 2.0.0        & 24.2K & 166 & 2.3K & 2 & 1 & 1 & 8 (0) & 0 & 1m51s & 16m \\
    JSON               & 2.4          & 28K & 172 & 5.9K & 3 & 2 & 1 & 9 (0) & 0& 1m52s & 18m \\
    JavassistWeld      & 3.12.1       & 60.4K & 813 & 11.3K & 2 & 1 & 1 & 8 (0) & 0 & 1m58s & 16m \\
    Jython             & 2.5.2        & 271.9K & 6.7K & 66.4K & 4 & 1 & 1 & 32 (1) & 0& 2m54s & 1h4m \\
    MozillaRhino       & 1.7R2        & 118.7K & 329 & 8.2K & 4 & 2 & 2 & 7 (2) & 2 & 1m56s & 12m10s   \\
    Myfaces            & 2.2.9        & 330.1K & 1.8K & 22.8K & 2 & 1& 2 & 7 (0) & 0& 2m1s & 14m   \\
    ROME               & 1.0          & 94.5K & 423 & 6.9K & 2 & 1 & 1 & 5 (1) & 1 & 1m48s & 8m53s  \\
    Spring             & 4.1.4        & 904.3K & 1.3K & 14.5K & 3 & 2 & 2 & 10 (0) & 0& 1m59s & 20m  \\
    Vaadin             & 7.7.14       & 572.1K & 4.5K & 17.5K & 4 & 1 & 1 & 13 (1) & 1& 1m54s & 24m37s  \\
    Wicket             & 6.23.0       & 420.7K & 3.2K & 11.1K & 2 & 1& 1 & 7 (0) & 0 & 1m50s & 14m \\
    \midrule
    \multicolumn{2}{c|}{\textbf{Total}} & - & - & - & - & - & 34 & 583 (20) & 16 & - & -\\
    \bottomrule
  \end{tabular}
  \label{EffectivenessResult}
\end{table*}
Table \ref{EffectivenessResult} summarizes the statistics about the evaluation results. The third to fifth columns present the scale of the target applications, including the lines of code (\emph{LoC}), the number of classes, and the number of methods. The sixth and seventh columns show the number of source methods and sink methods covered on an application's classpath. The eighth column represents the number of known gadget chains provided by ysoserial. The \emph{Identified Chains} and \emph{Confirmed Chains} columns respectively mean the number of gadget chains identified by \sysname's taint analysis module and the number of gadget chains reported by the fuzzing module. Note here that the number in parentheses of the \emph{Identified Chains} columns represents the number of truly exploitable gadget chains in the benchmark identified by the static identifier module, i.e., true positives (TP). For example, \sysname statically identified nine gadget chains in JDK, two of which are known gadget chains. In the dynamic validation process, these two chains were confirmed by \sysname with generated injection objects. The last two columns, \emph{Analysis Time} and \emph{Fuzzing Time}, show the total time overhead of taint analysis and fuzzing campaigns in each application.

Overall, in 22 Java libraries, \sysname statically identified a total of 20 out of 34 known gadget chains, and dynamically generated sink-reachable injection objects for 16 out of these chains without false positives. The results demonstrate the effectiveness of \sysname in discovering Java ODD gadget chains.

\noindent\textbf{False Positives.}
In the static analysis stage, we find that among the 583 identified gadget chains, \sysname correctly discovers 20 known gadget chains with a recall of 58.8\% (20/34). In other words, the false-positive rate (FPR) of \sysname in static analysis is 96.6\% (563/583). The root cause is mainly due to our simple static taint analysis logic. For example, given that some applications implement their own deserialization libraries/protocols (e.g., \code{XStream} \cite{XStream} and \code{Hessian} \cite{Hessian}) instead of using Java native deserialization interfaces (e.g., \code{Serializable} and \code{Externalizable}), \sysname takes all available methods (no matter whether the classes to which they belong inherit the \code{Serializable} or \code{Externalizable} interface) on the application's classpath into consideration for gadget chain construction, resulting in a large number of infeasible candidate gadget chains in practice. We discuss this limitation in Section \ref{Discussion} and leave the enhancement of static analysis as future effort. Considering that \sysname has validated these candidate gadget chains through structure-aware directed greybox fuzzing and confirmed 16 known gadget chains from 583 candidates with zero false positives, such a FPR is acceptable.

\noindent\textbf{False Negatives.}
As shown in Table \ref{EffectivenessResult}, we also find that 14 out of 34 (with a false-negative rate of 41.2\%) known gadget chains in ysoserial are missed by \sysname in static identification stage, mainly due to the limited support for certain dynamic features of Java language such as \emph{reflective calls} \cite{Reflection1} and \emph{dynamic proxy} \cite{DBLP:conf/issta/FourtounisKS18}. For example, in \code{Groovy1} \cite{Groovy1}, the attacker could exploit the class \code{ConvertedClosure}, whose constructor receives a proxy \code{MethodClosure} as its parameters, to pass tainted arguments to the gadget \code{MethodClosure.call()} to execute the malicious commands. Due to the unawareness to which classes can be proxied, gadget chains involving dynamic proxy during their construction are difficult to be identified by \sysname, resulting in false negatives.

In the dynamic verification stage, as shown in Table \ref{34Chains} in Appendix, there are four statically identified gadget chains (including \code{AspectJWeaver} \cite{AspectJWeaver}, \code{CommonsCollections1} \cite{CommonsCollections1}, \code{CommonsCollections3} \cite{CommonsCollections3}, and \code{Jython1} \cite{Jython1}) cannot be dynamically validated by \sysname. For \code{AspectJWeaver}, \sysname fails to generate sink-reachable injection objects because its sink method \code{writeToPath()} receives a file as input, which cannot be generated by traversing the property tree. For the remaining three gadget chains, their construction involves dynamic proxy\footnote{Although \sysname does not support for dynamic proxy, following GadgetInspector, we regard these known proxy classes as sources to start the gadget chain identification. Hence, these three dynamic proxy-related gadget chains can be statically identified by both \sysname and GadgetInspector.}, which is not supported by our injection object generation and mutation strategy.

\subsubsection{Impact of Structure-Aware Seed Generation}
\begin{figure}[t]
\centering
\begin{subfigure}{\linewidth}
\centering
\includegraphics[width=\linewidth]{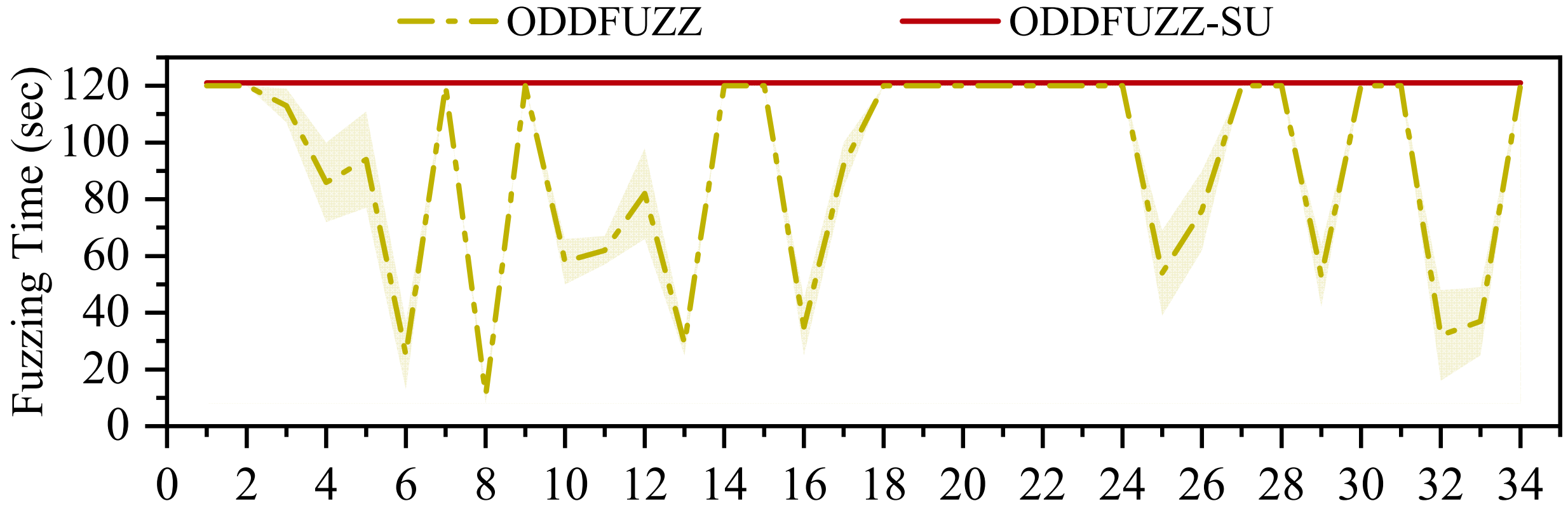}
%\quad
\end{subfigure}
\begin{subfigure}{\linewidth}
\centering
\includegraphics[width=\linewidth]{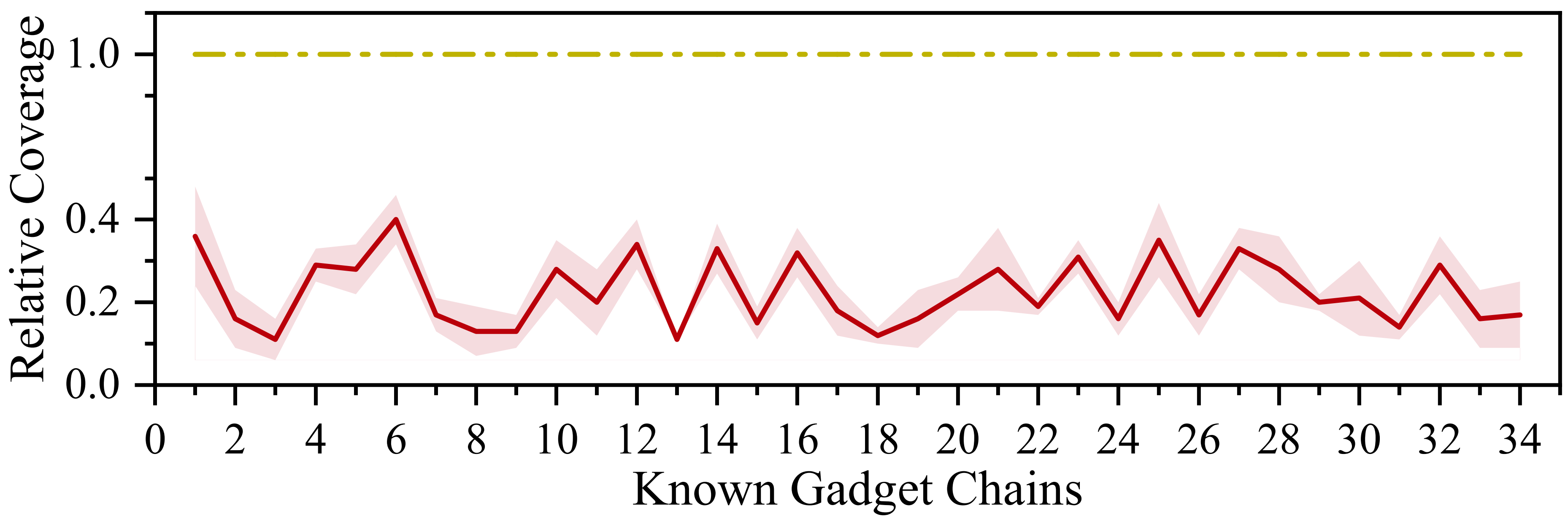}
\end{subfigure}
\caption{Comparison of \sysname and \sysname-SU. The $x$-axis is 34 known gadget chains listed in Table \ref{34Chains}. The $y$-axes are the fuzzing time and relative coverage (i.e., the ratio of gadget coverage that obtained by \sysname-SU and \sysname), respectively.}
\label{CHASER-SU}
\end{figure}

To construct valid seed objects for gadget chain fuzzing, we proposed a structure-aware seed approach, which leverages the class hierarchy relations between gadgets, to ensure both syntactic and semantic validity of inputs. To evaluate how structure-aware seed contributes to the gadget chain fuzzing of \sysname, we set up a naive variant of \sysname, \sysname-SU (SU: Structure-Unaware), which disables the structure-aware seed. We then reran the experiments 10 times.

The experimental results (the original gadget coverage is reported in Figure \ref{ConcreteCoverage1} in Appendix) are shown in Figure \ref{CHASER-SU}, where we can observe that our structure-aware input generator can effectively construct syntactically and semantically valid seed objects, successfully validating the target chain within the given time budget. In some cases (e.g., \code{CommonsCollections2} in \code{Apache Commons Collections4} library), \sysname takes only a dozen seconds to validate the target gadget chain. By contrast, \sysname-SU is unable to validate any reported gadget chain. That is because structure-unaware fuzzing has no prior knowledge of object structure provided by target gadget chains, thus stuck in the initial fuzzing stage. As shown in Figure \ref{CHASER-SU}, the lack of structured injection objects makes the fuzzer only trigger a few gadgets. This result demonstrates the effectiveness of structure-awareness in gadget chain fuzzing, which allows us to achieve performance improvement in dynamic validation.

\subsubsection{Impact of Feedback-Driven Fuzzing Guidance}
To effectively guide the injection objects to evolve towards desired gadgets, we proposed a feedback-driven fuzzing guidance that is armed with two key strategies, i.e., step-forward mutation and hybrid feedback-based seed prioritization. To evaluate how step-forward mutation and hybrid feedback contribute to the gadget chain fuzzing of \sysname, we also set up the variants of \sysname, \sysname-RM (RM: Random Mutation), \sysname-DG (DG: Distance-Guided) and \sysname-CG (CG: Coverage-Guided), which disables the step-forward mutation and decouples the distance feedback and coverage feedback, respectively. We then reran the experiments 10 times again.

\begin{figure}[t]
  \centering
  \includegraphics[width=\linewidth]{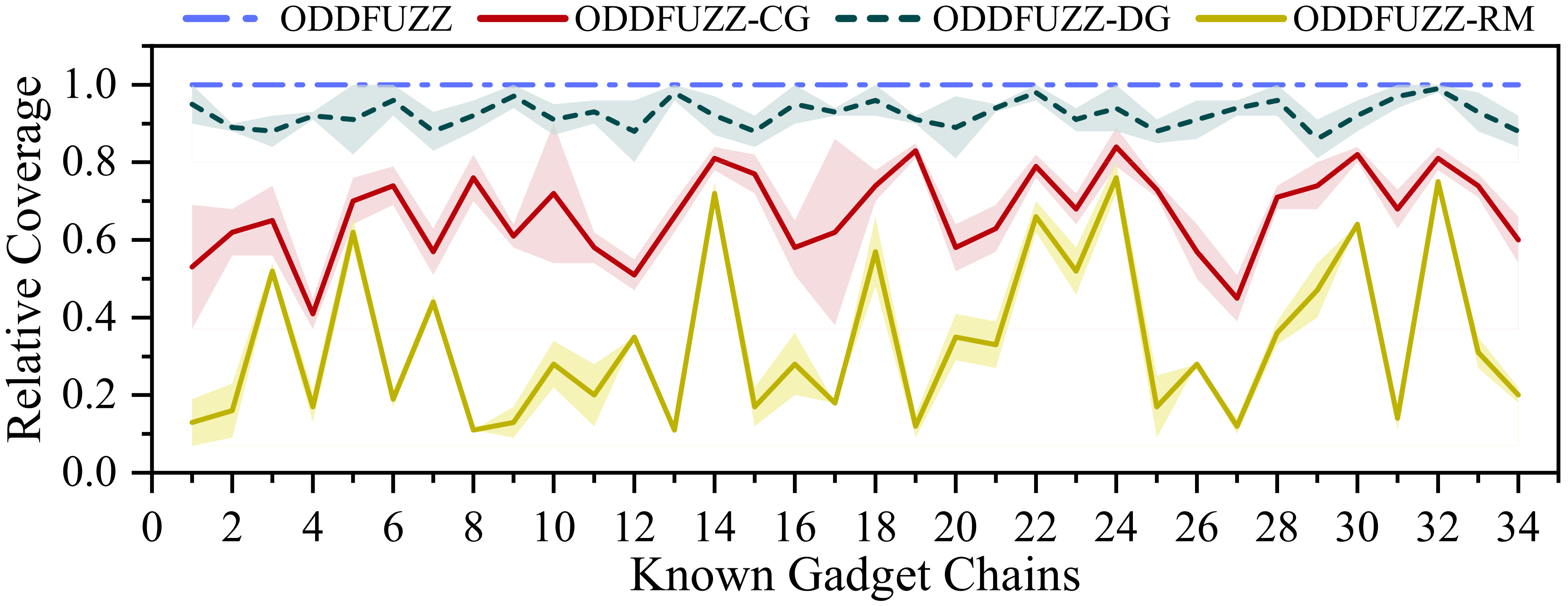}
\caption{Coverage comparison of \sysname, \sysname-RM, \sysname-CG and \sysname-DG. We use the results of \sysname as the baseline. The $x$-axis is 34 known gadget chains from ysoserial. The $y$-axis is the relative coverage of each variant against \sysname.}
\label{coverage}
\end{figure}

The experimental results (the original gadget coverage is reported in Figure \ref{ConcreteCoverage2} in Appendix) are shown in Figure \ref{coverage}, where we can observe that \sysname discovers more valid branches than \sysname-RM, \sysname-DG and \sysname-CG in almost every gadget chain, guiding the fuzzer to evolve towards desired gadgets. For instance, in 25 of 34 gadget chains (73.5\%), the random mutation-based fuzzer \sysname-RM only discovers branches that are less than half of \sysname. According to our analysis, it is mainly because blind mutation to object properties makes the fuzzer continue to explore shallow gadgets or infeasible paths, limiting the capability of the fuzzer to preserve critical waypoints during fuzzing. In addition, we also observe that compared to coverage-guided fuzzing, distance-guided fuzzing significantly increases the valid branches triggered by the fuzzer (the improvement is more than 40\% in certain cases). This result demonstrates the effectiveness of both strategies, step-forward mutation and hybrid feedback, as both of them contribute to the directedness of gadget chain fuzzing, and their combination allows \sysname to carry the fuzzing exploration towards the desired sinks.

\iffalse
\begin{tcolorbox}[left=1pt,right=1pt,top=1pt,bottom=1pt,boxrule=0.7pt,enhanced,drop fuzzy shadow]
\sysname confirms 17 exploitable gadget chains with zero false positives, owing to its structure-aware seed and feedback-driven fuzzing guidance in gadget chain fuzzing. By integrating two strategies together, the performance of gadget chain discovery can be significantly improved.
\end{tcolorbox}
\fi

\subsection{Comparison with State-of-the-Art Work}\label{ComparisonTools}

\iffalse
\begin{table}[t]\footnotesize
 \caption{Specification of baseline gadget chain detectors.}
  \centering
  \renewcommand\arraystretch{1.2}
  \renewcommand\tabcolsep{2pt}
  \begin{tabular}{cm{1.3cm}<{\centering}m{1.3cm}<{\centering}m{2cm}<{\centering}m{1.3cm}<{\centering}}
    \toprule
    \textbf{Technique} & \textbf{Static Analysis} & \textbf{Dynamic Analysis} & \textbf{Structure-Aware Inputs} & \textbf{Feedback-Driven}\\
    \midrule
    GadgetInspector   & \textcolor[RGB]{0,128,0}{\CheckmarkBold}  &  \textcolor[RGB]{209,26,66}{\usym{2717}}  & \textcolor[RGB]{209,26,66}{\usym{2717}} &   \textcolor[RGB]{209,26,66}{\usym{2717}} \\ 
    SerHybrid  & \textcolor[RGB]{0,128,0}{\CheckmarkBold}  &  \textcolor[RGB]{0,128,0}{\CheckmarkBold}   & \textcolor[RGB]{209,26,66}{\usym{2717}}  &   \textcolor[RGB]{209,26,66}{\usym{2717}} \\
    \textbf{\sysname}  & \textcolor[RGB]{0,128,0}{\CheckmarkBold}  & \textcolor[RGB]{0,128,0}{\CheckmarkBold} & \textcolor[RGB]{0,128,0}{\CheckmarkBold}   &  \textcolor[RGB]{0,128,0}{\CheckmarkBold} \\
    \bottomrule
  \end{tabular}
  \label{specification}
\end{table}
\fi

\begin{table*}[ht]\scriptsize
 \caption{Comparison with GadgetInspector and SerHybrid.}
  \centering
  \renewcommand\tabcolsep{2pt}
  \begin{tabular}{lc|m{1.37cm}<{\centering}m{1.37cm}<{\centering}m{1.1cm}<{\centering}|m{1.37cm}<{\centering}m{1.37cm}<{\centering}m{1.3cm}<{\centering}|m{1.37cm}<{\centering}m{1.37cm}<{\centering}m{1.1cm}<{\centering}m{1.16cm}<{\centering}}
    \toprule
    \multirow{3}*{\textbf{Application}} & \multirow{3}{0.95cm}{\textbf{Known Chains}}   & \multicolumn{3}{c|}{\textbf{\footnotesize GadgetInspector}} & \multicolumn{3}{c|}{\textbf{\footnotesize SerHybrid}} & \multicolumn{4}{c}{\textbf{\footnotesize \sysname}}\\
    ~ & ~ & Identified Chains  & Confirmed Chains  & Analysis Time & Identified Chains  & Confirmed Chains  & Analysis Time & Identified Chains  & Confirmed Chains  & Analysis Time & Fuzzing Time\\
    \midrule
    JDK                & 4 & 5 & 0 & 53s & N/A & N/A & N/A & 9 (1) & 1 & 1m51s & 16m32s \\
    AspectJWeaver      & 1 & 6 & 0 & 41s & N/A & N/A & N/A & 9 (1) & 0 & 1m56s & 18m \\ 
    BeanShell          & 1 & 2 & 0 & 49s & 1 & 0 & 10m55s & 8 (0) & 0 & 1m53s & 16m \\ 
    C3P0               & 1 & 2 & 0 & 48s & N/A & N/A & N/A & 13 (1) & 1 & 1m50s &  25m53s \\ 
    Click              & 1 & 4 & 0 & 39s & N/A & N/A & N/A & 8 (1) & 1 & 1m48s & 15m26s \\ 
    Clojure            & 1 & 12 & 1 & 40s & N/A & N/A & Timeout & 184 (1) & 1 & 3m30s & 6h7m34s \\ 
    CommonsBeanutils   & 1 & 2 & 0 & 37s & 0 & 0 & 13m6s & 8 (1) & 1 & 1m52s & 14m25s \\ 
    CommonsCollections & 5 & 4 & 1 & 39s & 1 & 1 & 26m51s & 97 (5) & 3 & 1m58s & 3h10m53s\\ 
    CommonsCollections4& 2 & 4 & 0 & 38s & 1 & 1 & 11m21s & 112 (2) & 2 & 1m55s & 3h41m9s\\ 
    FileUpload         & 1 & 3 & 0 & 38s & N/A & N/A & N/A & 8 (0) & 0 & 1m55s & 16m \\ 
    Groovy             & 1 & 4 & 0 & 47s & 3 & 0 & 1h26m & 13 (0) & 0 & 2m8s & 26m \\ 
    Hibernate          & 2 & 3 & 0 & 41s & 3 & 0 & 56m37s & 8 (2) & 2 & 2m8s & 14m7s \\ 
    JBossInterceptors  & 1 & 2 & 0 & 38s & N/A & N/A & N/A & 8 (0) & 0 & 1m51s & 16m \\ 
    JSON               & 1 & 2 & 0 & 39s & N/A & N/A & N/A & 9 (0) & 0 & 1m52s & 18m\\ 
    JavassistWeld      & 1 & 2 & 0 & 39s & N/A & N/A & N/A & 8 (0) & 0 & 1m58s & 16m \\ 
    Jython             & 1 & 42 & 1 & 50s & N/A & N/A & Timeout & 32 (1) & 0 & 2m54s &  1h4m\\ 
    MozillaRhino       & 2 & 3 & 0 & 40s & N/A & N/A & N/A & 7 (2) & 2 & 1m56s & 12m10s \\ 
    Myfaces            & 2 & 2 & 0 & 37s & N/A & N/A & N/A & 7 (0) & 0 & 2m1s & 14m \\ 
    ROME               & 1 & 2 & 0 & 36s & 0 & 0 & 6m30s & 5 (1) & 1 & 1m48s & 8m53s \\ 
    Spring             & 2 & 2 & 0 & 38s & N/A & N/A & N/A & 10 (0) & 0 & 1m59s & 20m\\ 
    Vaadin             & 1 & 5 & 0 & 37s & N/A & N/A & N/A & 13 (1) & 1 & 1m54s & 24m37s \\ 
    Wicket             & 1 & 3 & 0 & 36s & N/A & N/A & N/A & 7 (0) & 0 & 1m50s & 14m \\ 
    \midrule
    \textbf{Total}     & 34 & 116 & 3 & - & 9 & 2 & - & 583 (20) & 16 & - & - \\
    \bottomrule
  \end{tabular}
  \label{ComparisonResult}
\end{table*}

We compared \sysname with two state-of-the-art automated gadget chain discovery tools, GadgetInspector \cite{Inspector} and SerHybrid \cite{Serhybrid}. As shown in Table \ref{ComparisonResult} (detailed results\footnote{Unfortunately, despite our best effort, the implementation of SerHybrid was not reproducible. We made unsuccessful attempts to contact the authors for suggestions. Hence, we compare them against the results published in their paper. To ensure fairness, we carefully make the experimental settings and only compare them with the same libraries they tested.} are reported in Table \ref{34Chains} in Appendix). The \emph{Identified Chains} column represents the number of potential gadget chains statically identified by each approach, and the \emph{Confirmed Chains} column indicates how many of them are disclosed as exploitable in ysoserial. Considering that GadgetInspector is purely static, its \emph{Confirmed Chains} column denotes the static results, while in SerHybrid and \sysname, the \emph{Confirmed Chains} column shows the results after dynamic validation. Similar to Table \ref{EffectivenessResult}, the \emph{Analysis Time} and \emph{Fuzzing Time} columns respectively represent the total time cost of static analysis and fuzzing campaigns in each approach. Note that, as reported in \cite{Serhybrid}, 13 out of 22 applications (involving 19 exploitable gadget chains) are not evaluated by SerHybrid because it specifically focuses on reflection-enabled Java ODD vulnerabilities (labeled as ``N/A"), and two applications (\code{Clojure} and \code{Jython}) cannot be analyzed statically within given time budgets (labeled as ``Timeout").

Overall, \sysname achieves significant performance improvement in all applications. In particular, \sysname reported 16 out of 34 exploitable gadget chains without false positives, including 13 unique gadget chains that cannot be found by baselines. By contrast, the number of truly exploitable gadget chains reported by GadgetInspector and SerHybrid is three and two, respectively.

\noindent\textbf{\sysname vs. GadgetInspector.}
As shown in Table \ref{ComparisonResult}, GadgetInspector takes an average of 41 seconds to analyze each application and reports 116 suspicious gadget chains. However, only three of them are exploitable, meaning that 97.4\% of them are false positives. Such a significant performance gap mainly results from two aspects. On the one hand, constrained by a few simplifying assumptions (e.g., all members of a tainted object are also tainted) and requiring manual inspection of the reports, GadgetInspector is prone to precision issues and cannot guarantee that identified gadget chains are truly exploitable. On the other hand, due to the lack of consideration of Java runtime polymorphism when computing intraprocedural data flows, GadgetInspector suffers from unsound analysis results, resulting in missed available gadgets, i.e., false negatives. By contrast, as reported in Table \ref{34Chains}, benefiting from our lightweight summary-based taint analysis, \sysname statically identifies 17 more exploitable gadget chains (covering all three chains reported by GadgetInspector) and dynamically validated 15 out of them with no false positives. It is noteworthy that \code{CommonsCollections1} \cite{CommonsCollections1}, which can be identified by GadgetInspector, fails to be validated by our approach because of certain specific cases (e.g., dynamic proxy discussed in Section \ref{Effectiveness}) in injection object construction. Nevertheless, the detection capability of \sysname is still promising (significantly improving the recall rate of static analysis with acceptable time overhead) and these limitations can be solved to some extent (as discussed in Section \ref{Discussion}).

\noindent\textbf{\sysname vs. SerHybrid.}
As reported in Table \ref{ComparisonResult}, SerHybrid successfully confirms two exploitable gadget chains in under two minutes on average. Despite its promising performance, a major drawback of SerHybrid lies in that most (32 out of 34) exploitable gadget chains are missed.For instance, although SerHybrid statically identifies three potential gadget chains in \code{Hibernate} in under one hour, it cannot generate an injection object for any of them for validation within 30 minutes (the time budget set by \cite{Serhybrid}). According to our manual analysis, it is mainly because an execution path from a source object to the sink object with a tainted flow cannot provide the fuzzer with class hierarchy information required for injection object generation. By contrast, owing to our structure-aware directed greybox fuzzing, \sysname efficiently generates both syntactically and semantically valid injection objects for 16 exploitable gadget chains (including two chains in \code{Hibernate}) within two minutes on average.

\iffalse
\begin{tcolorbox}[left=1pt,right=1pt,top=1pt,bottom=1pt,boxrule=0.7pt,enhanced,drop fuzzy shadow]
\sysname significantly outperforms the state-of-the-art automated Java deserialization gadget chain discovery tools. Furthermore, \sysname can identify 14 unique gadget chains that cannot be found by other tools.
\end{tcolorbox}
\fi

\subsection{Vulnerability Discovery}\label{VulnerabilityHunting}

We chose target applications that satisfied the following criteria. First, they are Java projects since \sysname is designed to search deserialization vulnerabilities in Java. Second, as discussed in our threat model, these applications should contain known deserialization entries because we prefer to actually exploit the Java deserialization vulnerabilities with found gadget chains rather than just discover potential chains. Third, they cover diverse application domains so that the generality of our approach can be evaluated. Using the three criteria, we selected four target Java applications, including \code{Oracle WebLogic Server} (a commercial application server), \code{Sonatype Nexus} (a repository manager), \code{Apache Dubbo} (a high-performance Remote Procedure Call (RPC) framework), and \code{protostuff} (a Java serialization library), to demonstrate the vulnerability discovery capability of \sysname in practical scenarios.

\subsubsection{Unknown Vulnerability Discovery}

An overview of the vulnerabilities found by \sysname is shown in Table \ref{hunting} in Appendix. In total, \sysname has successfully detected six previously unknown Java ODD vulnerabilities. While three of them were found within \code{Oracle WebLogic Server}, the remaining three vulnerabilities respectively arose from \code{Sonatype Nexus}, \code{Apache Dubbo}, and \code{protostuff}. These vulnerabilities can be exploited to perform RCE attacks by building our newly discovered gadget chains. We have responsibly reported all the vulnerabilities to corresponding vendors and have received their positive feedback. At the time of paper writing, five of the vulnerabilities have been patched and assigned CVE numbers due to their severe security consequences.

\subsubsection{Case Study}
\begin{figure}[ht]
\centering
\begin{subfigure}{\linewidth}
\centering
\includegraphics[width=\linewidth]{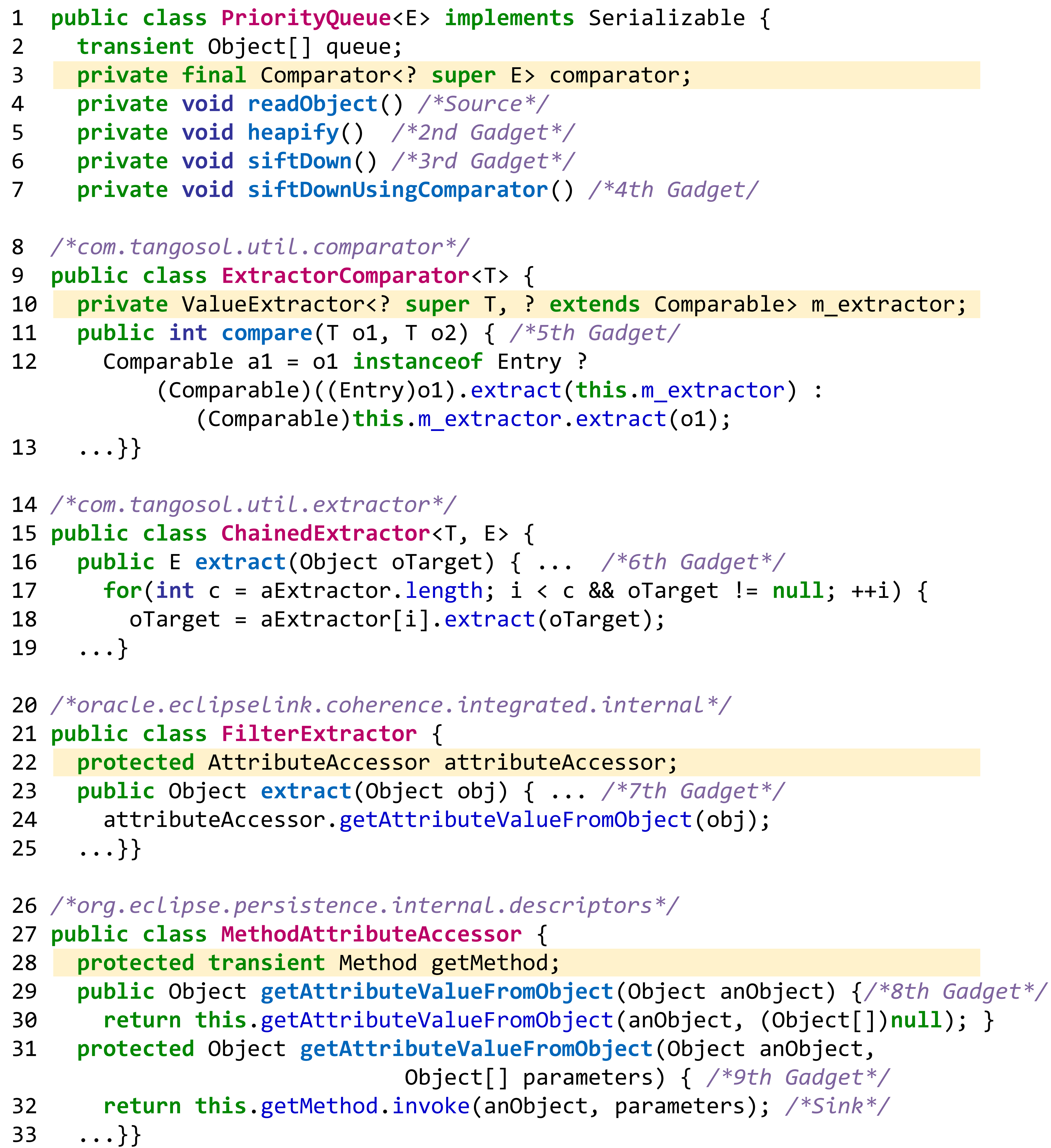}
\caption{\label{CaseCode}A simplified code snippet of the gadget chain for CVE-2020-14756 in WebLogic.}
\quad
\end{subfigure}
\begin{subfigure}{\linewidth}
\centering
\includegraphics[width=\linewidth]{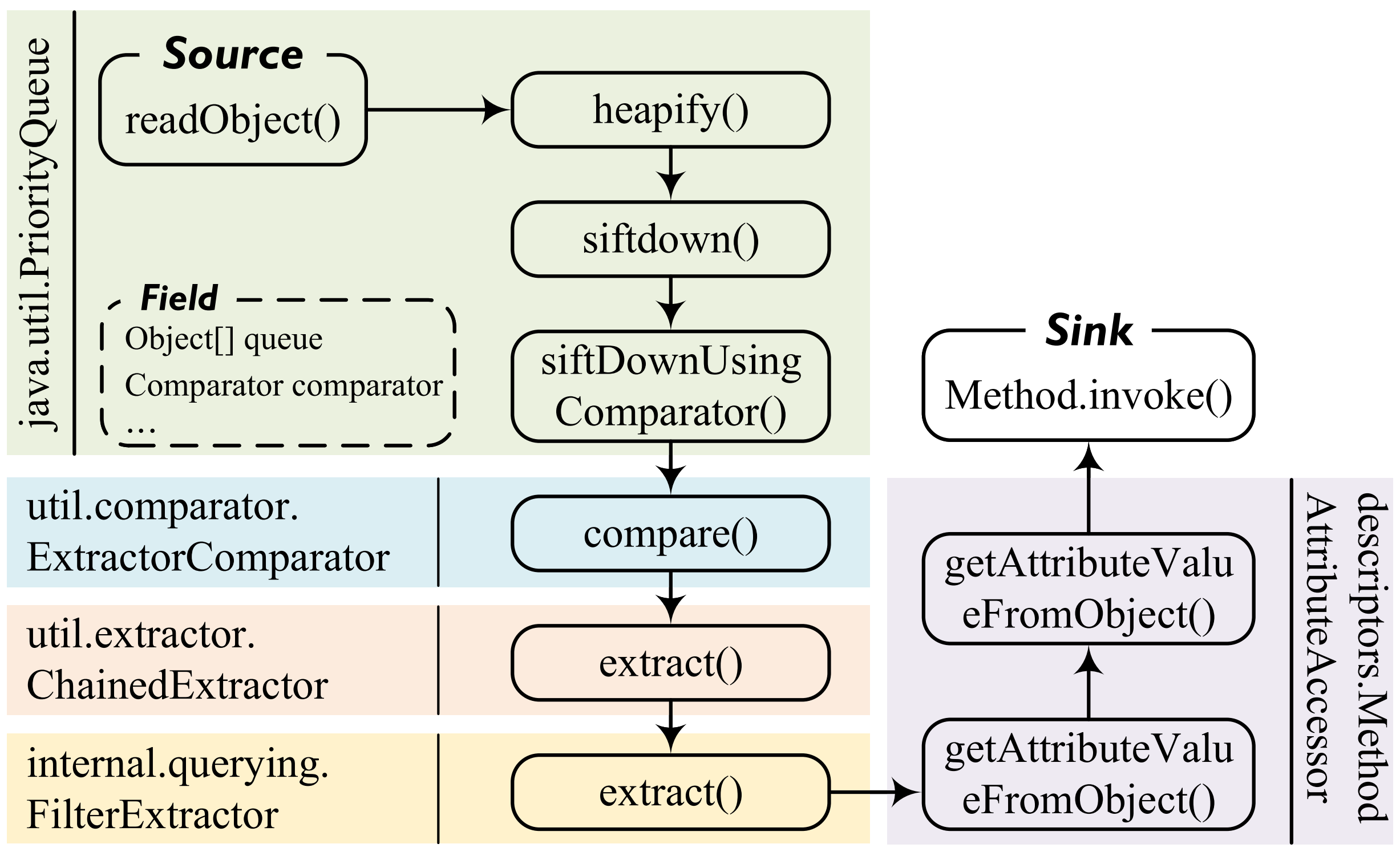}
\caption{\label{CaseChain} An exploitable gadget chain for \ref{CaseCode}.}
\end{subfigure}
\caption{\label{CaseStudy}A previously unknown vulnerability found by \sysname.}
\end{figure}

\sysname uncovered a RCE vulnerability (CVE-2020-14756 \cite{CVE-2020-14756}) in the \code{Oracle Coherence} product of \code{Oracle WebLogic Server}. Successful attack of this vulnerability can result in takeover of \code{Oracle Coherence}. As shown in Figure \ref{CaseCode}, the flow of triggering the vulnerability is as follows: \ding{182} The attacker first instantiates \code{PriorityQueue} to reuse the known entry gadget (magic method) \code{readObject()} which unconditionally invokes the second to fourth gadget (line 5-7). \ding{183} To connect the fifth gadget \code{ExtractComparator.compare()}, the field \code{comparator} of class \code{PriorityQueue} should be set to the \code{ExtractorComparator}'s instance through POP. \ding{184} Following the above steps to recursively modify the injection object's properties to trigger the security-sensitive call site \code{Method.invoke()} (line 32). The complete gadget chain of CVE-2020-14756 is shown in Figure \ref{CaseChain} in Appendix.

It is difficult to validate this exploitable gadget chain by traditional fuzzing, as it requires prior knowledge about the layout of the multilevel sub-objects to avoid the fuzzing campaign stuck in the initial stage due to randomly generated property values. However, \sysname is able to produce a syntactically valid injection object to facilitate gadget chain execution by supplying the fuzzed structure extracted from the nested hierarchy of multiple gadget classes to the input generator. In this way, the injection object can be fuzzed to reach the sink. Moreover, in the fuzzing campaign, \sysname mutated the fuzzed object step by step towards the sink and finally triggered the vulnerability.

\section{Discussion}\label{Discussion}
\noindent\textbf{Better Static Analysis.}
As discussed before, our static analysis suffers from precision and soundness problems. Specifically, false positives can be introduced by our simple taint analysis logic. For example, \sysname does not restrict the candidate space of available gadgets according to application-specific deserialization libraries/interfaces and not handle several special cases (e.g., an untrusted variable modified by the keyword \code{transient} cannot be deserialized). These false positives can be reduced by improving taint analysis rules. By contrast, false negatives can be introduced by missing indirect call targets caused by certain dynamic features (reflective calls, dynamic proxy, etc.). Fortunately, benefiting from recent solutions \cite{Reflection1,DBLP:conf/issta/FourtounisKS18} which can (partially) solve these advanced language features, the unsoundness of our approach can be well mitigated. In addition, similar to existing works \cite{Inspector,Serhybrid,FUGIO,DBLP:conf/ccs/DahseKH14}, the effectiveness (recall) of our static gadget chain identification also relies heavily on the prior expert knowledge of available sources and sinks, which is the main manual effort required in \sysname. Considering that there are a few orthogonal tools/approaches \cite{Scanner,objectMap} have been proposed to automatically identify untrusted deserialization entry points, and our knowledge base is configurable, i.e., newly disclosed sources and sinks can be dynamically added, the capability to detect unknown Java ODD vulnerabilities in the wild can be improved.

\noindent\textbf{Diverse Generation Strategy.}
As evaluated in Section \ref{Effectiveness}, the awareness of object structure can help improve the performance in gadget chain fuzzing. However, the optimal generation strategy needs to be suitable for diverse deserialization scenarios and might be changed depending on the construction of a gadget chain. For example, the exploitation of certain available gadgets relies on some specific techniques (e.g., the dynamic proxy used in \code{Groovy1} \cite{Groovy1}) or constraints (e.g., the file input required by \code{AspectJWeaver} \cite{AspectJWeaver}), which blocks the injection object generation with the property tree. A possible solution is to design some general templates for these specifications. We leave such exploration for our future work.

\noindent\textbf{Exploit Construction.}
\sysname also requires some human efforts to help construct practical exploits because the injection objects constructed by our structure-aware generator represent the minimum effort required to trigger the gadget chains. In order to construct actual exploits, security analysts need to further \ding{182} check whether the tainted properties flowing into the security-sensitive call site are attacker-controllable. If it is truly attacker-controllable, security analysts should \ding{183} manually replace the non-harmful command (e.g., open calculator) with a malicious one (e.g., reverse shell) based on the injection objects generated by \sysname. We intend to find a more automatic way as future work, while in this work, the core goal of \sysname is to efficiently discover exploitable Java ODD gadget chains from a large number of static analysis reports.

\section{Related Work}
\subsection{Deserialization Vulnerabilities in Java}

\noindent\textbf{Vulnerability Mitigation.}
Most existing works focus on understanding and protecting applications against known deserialization vulnerabilities \cite{DBLP:conf/uss/AzadLN19}. Muñoz et al. \cite{Munoz} conducted a comprehensive analysis on JSON deserialization libraries and presented several mitigation measures as takeaways. Carettoni \cite{Carettoni} presented a configurable Java deserialization library, which supports multiple optional settings such as blacklist and whitelist, to secure application from untrusted input. Cristalli et al. \cite{DBLP:conf/raid/CristalliVBL18} designed a novel sandbox system, which collects the behavior information of benign deserialization process and constructs the precise execution path, to mitigate the problems of deserialization of untrusted data in Java.

\noindent\textbf{Vulnerability Detection.}
Despite the existing efforts, these defense solutions will be bypassed once a new gadget or fundamental vector is found \cite{blackhat2019}. Hence, some works focus on detecting potential vulnerabilities in applications \cite{DBLP:journals/infsof/CaoSBWL21,DBLP:journals/iet-sen/SubhanWBSR22,DBLP:journals/infsof/ZhouSXLC19}. Koutroumpouchos et al. \cite{objectMap} proposed an extendable tool \emph{ObjectMap} which generates a series of requests to detect whether the payload can be directly passed to the target application. Similarly, Marshalsec \cite{marshalsec} and Java Deserialization Scanner \cite{Scanner} are two tools that dynamically scan and exploit know gadget chains from the ysoserial project \cite{ysoserial}.

\noindent\textbf{Automated Gadget Chain Discovery.}
In order to automatically identify new gadget chains, Haken \cite{Inspector} presented GadgetInspector, which leverages static taint analysis and simple symbolic execution to mine the propagation paths of parameters within/between methods of a target application, and then performs a breadth-first search (BFS) to search for attacker-controllable gadget chains. Chen et al. \cite{Tabby} developed Tabby, a graph-based static analysis tool, to support gadget chain discovery. Rasheed et al. \cite{Serhybrid} proposed SerHybrid, a hybrid analysis-based approach which constructs a heap abstraction to produce actual injection objects to automatically validate exploitable gadget chains. Cao et al. \cite{GCMiner} proposed GCMiner, which captures both explicit and implicit method calls to identify candidate gadget chains, and adopts an overriding-guided object generation approach to guarantee the validity of injection objects during fuzzing. By contrast, \sysname aims to improve the effectiveness and efficiency of gadget chain validation via structure-aware directed greybox fuzzing.

\subsection{Deserialization Vulnerabilities in Other Languages}
Security threats of insecure deserialization also exist in other mainstream programming languages \cite{DBLP:conf/ccs/DahseKH14,javascript,.Net}. Dahse et al. \cite{DBLP:conf/ccs/DahseKH14,DBLP:conf/ndss/DahseH14} conducted static taint analysis to detect gadget chains in common PHP applications. FUGIO \cite{FUGIO} combined coarse-grained program analysis and fuzzing to automatically produce exploit objects for PHP Object Injection (POI) vulnerabilities. Shahriar and Haddad \cite{DBLP:conf/sac/ShahriarH16} proposed a lightweight approach based on latent semantic indexing to identify Object Injection Vulnerabilities (OIVs) in web application. They identified multiple keywords that are likely responsible for OIVs and defined customized queries to identify relevant source files to discover new vulnerabilities. SerialDetector \cite{.Net} studied the root cause of OIVs in .NET applications and presented a scalable taint-based data flow analysis to discover and leverage publicly available gadgets.

\section{Conclusion}

In this paper, we propose a novel hybrid solution \sysname to efficiently discover Java deserialization vulnerabilities. \sysname performs lightweight static taint analysis to identify candidate \emph{gadget chains} and applies a structure-aware directed fuzzing to mitigate false positives. Results show that, \sysname could discover 16 out of 34 known gadget chains, while two state-of-the-art baselines only identify three of them. Moreover, we have discovered six previously unreported exploitable gadget chains and five of them have been assigned with CVE-IDs.

\section*{Acknowledgment}

This work is supported in part by the National Natural Science Foundation of China (No.61972335, No.62202414, No.62002309, No.61972224); the CCF-AFSG Research Fund (No.CCF-AFSG RF20210022); the Six Talent Peaks Project in Jiangsu Province (No. RJFW-053), the Jiangsu ``333'' Project; the Open Funds of State Key Laboratory for Novel Software Technology of Nanjing University (No.KFKT2022B17), Yangzhou University Top-level Talents Support Program (2019).

\bibliographystyle{./bibliography/IEEEtran}
\bibliography{./bibliography/IEEEabrv,./bibliography/IEEEexample}

\appendices
\section{Target Sources and Sinks} \label{Source&Sinks}

Magic methods (sources) and security-sensitive call sites (sinks) covered by \sysname are listed below, among which six sources and 16 sinks are considered by GadgetInspector (highlighted in gray). These sensitive call sites can be exploited to perform \emph{Remote Code Execution} (RCE), \emph{JNDI injection} (JNDIi), \emph{System Resource Access} (SRA), and \emph{Server-Side Request Forgery} (SSRF) attacks. It is noteworthy that, similar to GadgetInspector, we explicitly maintain a set of magic methods (security-sensitive call sites) of specific classes as sources (or sinks) for gadget chain identification because not all of them can be exploited (e.g., in \code{Clojure}, only \code{clojure.main\$eval\_opt.invoke()} instead of \code{clojure.core\$comp\$fn\_4727.invoke()} is vulnerable).

\begin{itemize}[leftmargin=1em]
    \item \textbf{Magic Methods: }\code{\sethlcolor{gray}{\hl{readObject}}}, \code{\sethlcolor{gray}{\hl{hashCode}}}, \code{get}, \code{put}, \code{compare}, \code{readExternal}, \code{readResolve}, \code{\sethlcolor{gray}{\hl{finalize}}}, \code{\sethlcolor{gray}{\hl{equals}}}, \code{compareTo}, \code{toString}, \code{validateObject}, \code{readObjectNoData}, \code{<clinit>}, \code{\sethlcolor{gray}{\hl{call}}}, \code{\sethlcolor{gray}{\hl{doCall}}}
    \item \textbf{Security-Sensitive Call Sites:}
    
    \textbf{- \emph{Remote Code Execution} (RCE): }\code{getDeclaredMethod}, \code{getConstructor}, \code{findClass}, \code{\sethlcolor{gray}{\hl{getMethod}}}, \code{loadClass}, \code{start}, \code{\sethlcolor{gray}{\hl{exec}}}, \code{\sethlcolor{gray}{\hl{invoke}}}, \code{\sethlcolor{gray}{\hl{forName}}}, \code{\sethlcolor{gray}{\hl{newInstance}}}, \code{\sethlcolor{gray}{\hl{exit}}}, \code{defineClass}, \code{\sethlcolor{gray}{\hl{call}}}, \code{\sethlcolor{gray}{\hl{invokeMethod}}}, \code{\sethlcolor{gray}{\hl{invokeStaticMethod}}}, \code{\sethlcolor{gray}{\hl{invokeConstructor}}}

    \textbf{- \emph{JNDI Injection} (JNDIi): }\code{getConnection}, \code{do\_lookup}, \code{lookup}, \code{c\_lookup}, \code{getObjectInstance}, \code{connect}

    \textbf{- \emph{System Resource Access} (SRA): }\code{\sethlcolor{gray}{\hl{newBufferedReader}}}, \code{\sethlcolor{gray}{\hl{newBufferedWriter}}},
    \code{delete}, \code{\sethlcolor{gray}{\hl{newInputStream}}}, \code{\sethlcolor{gray}{\hl{newOutputStream}}}, \code{\sethlcolor{gray}{\hl{<init>}}}

    \textbf{- \emph{Server-Side Request Forgery} (SSRF): }\code{openConnection}, \code{\sethlcolor{gray}{\hl{openStream}}}
\end{itemize}

\section{Main fuzzing loop of \sysname} \label{AlgorithmDetails}

\begin{algorithm}
\footnotesize
	\caption{Gadget Chain Fuzzing}
	\label{algorithm}
	\begin{algorithmic}[1]
		\Require 
		the chain to be validated \emph{c}
		\Ensure
		sink-triggering seed set $\mathcal{S}_{\textcolor[RGB]{209,26,66}{\usym{2717}}}$
		\State $\mathcal{S}_{\textcolor[RGB]{209,26,66}{\usym{2717}}} \leftarrow \emptyset$ \label{emptyset}
		\State $\mathcal{S} \leftarrow$ {\scriptsize GENERATE}I{\scriptsize NITIAL}S{\scriptsize EED}(\emph{c}) \label{GenerateInitialSeed}
		\State \emph{minDistance} $\leftarrow \infty$ \label{distanceInit}
		\State \emph{gadgetCoverage} $\leftarrow \emptyset$ \label{coverageInit}
		\Repeat
		\State \emph{s} $\leftarrow$ {\scriptsize SELECT}S{\scriptsize EED}($\mathcal{S}$) \label{selectSeed}
		\State \emph{p} $\leftarrow$ {\scriptsize ASSIGN}E{\scriptsize NERGY}(\emph{s}) \label{assignEnergy}
		\For{\emph{i} from $1$ to \emph{p}}
		    \State \emph{s$'$} $\leftarrow$ {\scriptsize MUTATE}S{\scriptsize EED}(\emph{s}) \label{mutateSeed}
		    \State \emph{result} $\leftarrow$ {\scriptsize EXECUTE}P{\scriptsize ROGRAM}(\emph{s$'$}) \label{executeProgram}
		    \If{{\scriptsize REACH}S{\scriptsize INK}(\emph{result}) == \emph{true}} \label{reachSink}
		         \State $\mathcal{S}_{\textcolor[RGB]{209,26,66}{\usym{2717}}} \leftarrow \mathcal{S}_{\textcolor[RGB]{209,26,66}{\usym{2717}}}$ $\cup$ \emph{s$'$}
		         \State {\scriptsize EMIT}S{\scriptsize IGNAL}(\emph{c}, $``$\emph{Reachable}$"$) \label{signal}
		    \ElsIf{\emph{s$'$.distance} $<$ \emph{minDistance}} \label{mindistance}
		         \State $\mathcal{S} \leftarrow \mathcal{S}$ $\cup$ \emph{s$'$} 
		         \State \emph{minDistance} $\leftarrow$ \emph{s$'$.distance} \label{updateDistance}
		    \ElsIf{\emph{s$'$.coverage} $\nsubseteq$ \emph{gadgetCoverage}} \label{maxCoverage}
		         \State $\mathcal{S} \leftarrow \mathcal{S}$ $\cup$ \emph{s$'$}
		         \State \emph{gadgetCoverage} $\leftarrow$ \emph{s$'$.coverage} \label{updateCoverage}
		    \EndIf
		\EndFor
		\Until \emph{timeout} or \emph{sink-triggering} signal received \label{fuzzstop}
	\end{algorithmic} 
\end{algorithm}

Algorithm \ref{algorithm} describes the overall process of our gadget chain fuzzing. Given a target Java application and an identified gadget chain, the fuzzer initiates a fuzzing campaign during a given time budget. The fuzzing process starts by adding the initial injection object generated from the candidate gadget chain $c$ to the prepared seed pool $S$ (line \ref{emptyset}-\ref{GenerateInitialSeed}) and initializing feedback information (line \ref{distanceInit}-\ref{coverageInit}). It then repeats the following fuzzing loops until finding an injection object that can reach the security-sensitive call site in the gadget chain. To schedule favored seeds for fuzzing in each round, the fuzzer selects a set of seeds $s$ with higher priority from the seed pool $S$ based on their previous execution feedback (line \ref{selectSeed}). Each chosen seed is assigned to a certain amount of power that determines how many new seed inputs can be derived in this round (line \ref{assignEnergy}). Next, the fuzzer mutates the scheduled seed to generate a new injection object $s'$ to execute the instrumented program (line \ref{mutateSeed}-\ref{executeProgram}). When this mutated seed $s'$ reaches the security-sensitive call site, the fuzzer adds this seed to the sink-triggering seed set $\mathcal{S}_{\textcolor[RGB]{209,26,66}{\usym{2717}}}$ and emits a signal to \sysname to stop the fuzzing campaign of the target gadget chain (line \ref{reachSink}-\ref{signal}). If this mutated seed does not reach the sink but contributes to reducing the seed distance towards the target sink, the fuzzer derives new seeds and updates the current minimal seed distance \emph{minDistance} (line \ref{mindistance}-\ref{updateDistance}). Furthermore, if this mutated seed executes more branches within gadgets on the execution path of the target chain, the fuzzer also adds this seed to the seed pool and updates the current gadget coverage \emph{gadgetCoverage} (line \ref{maxCoverage}-\ref{updateCoverage}). The fuzzing loop will not stop until the given time budget expires or sink-triggering signal is received by \sysname (line \ref{fuzzstop}). The remaining section details each step in the fuzzing process.

\section{Hyperparameter Evaluation} \label{Sensitivity}
To evaluate the optimal setting of two hyperparameters, including the maximum length of gadget chains (\emph{Gadget Chain Length}) that \sysname analyzes and the time budget assigned to fuzzing each candidate gadget chain (\emph{Fuzzing Time Budget}), we conducted the following experiments.

\begin{figure}[ht]
\centering
\begin{subfigure}{\linewidth}
\centering
\includegraphics[width=\linewidth]{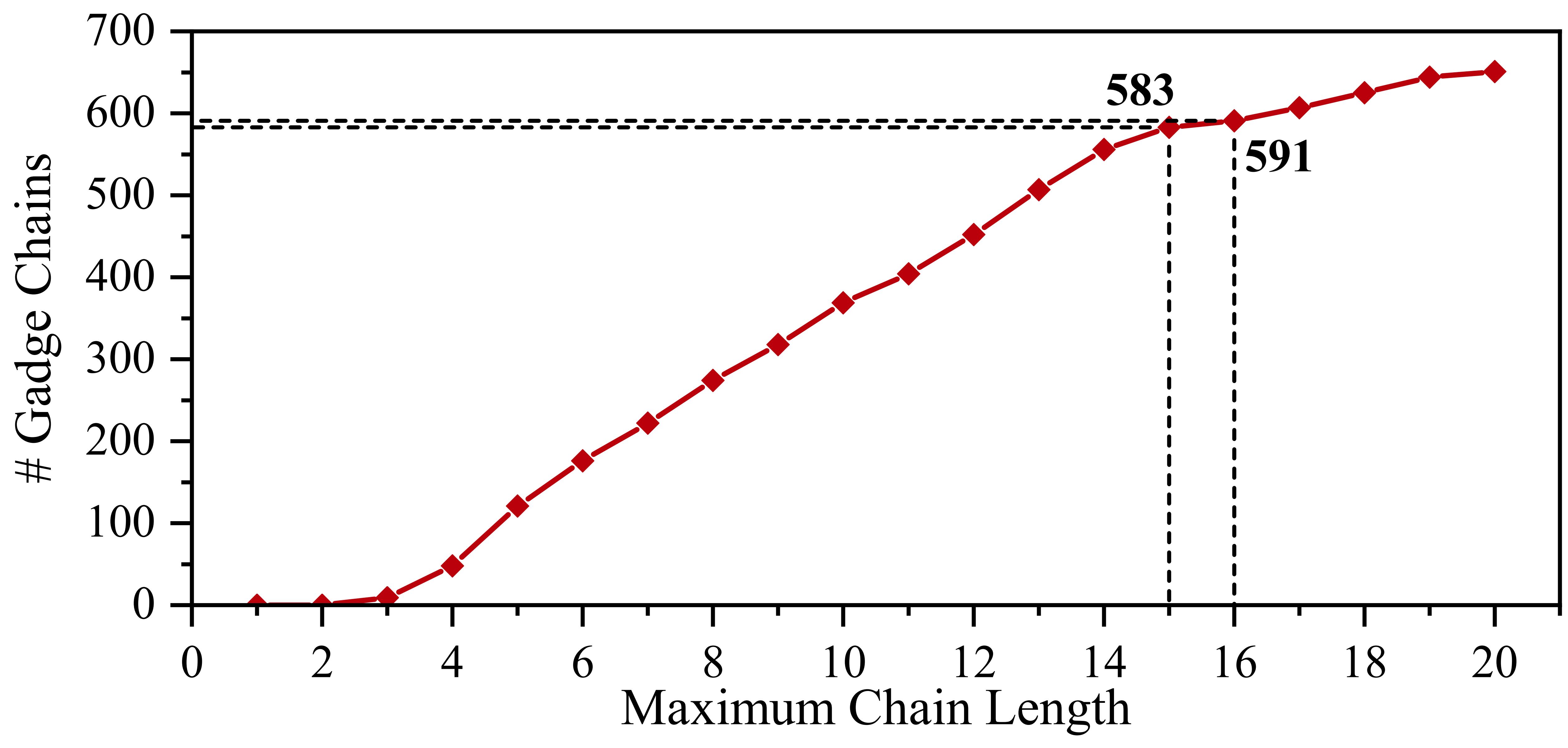}
\caption{\label{Length}Gadget chain length.}
\end{subfigure}
\begin{subfigure}{\linewidth}
\centering
\includegraphics[width=\linewidth]{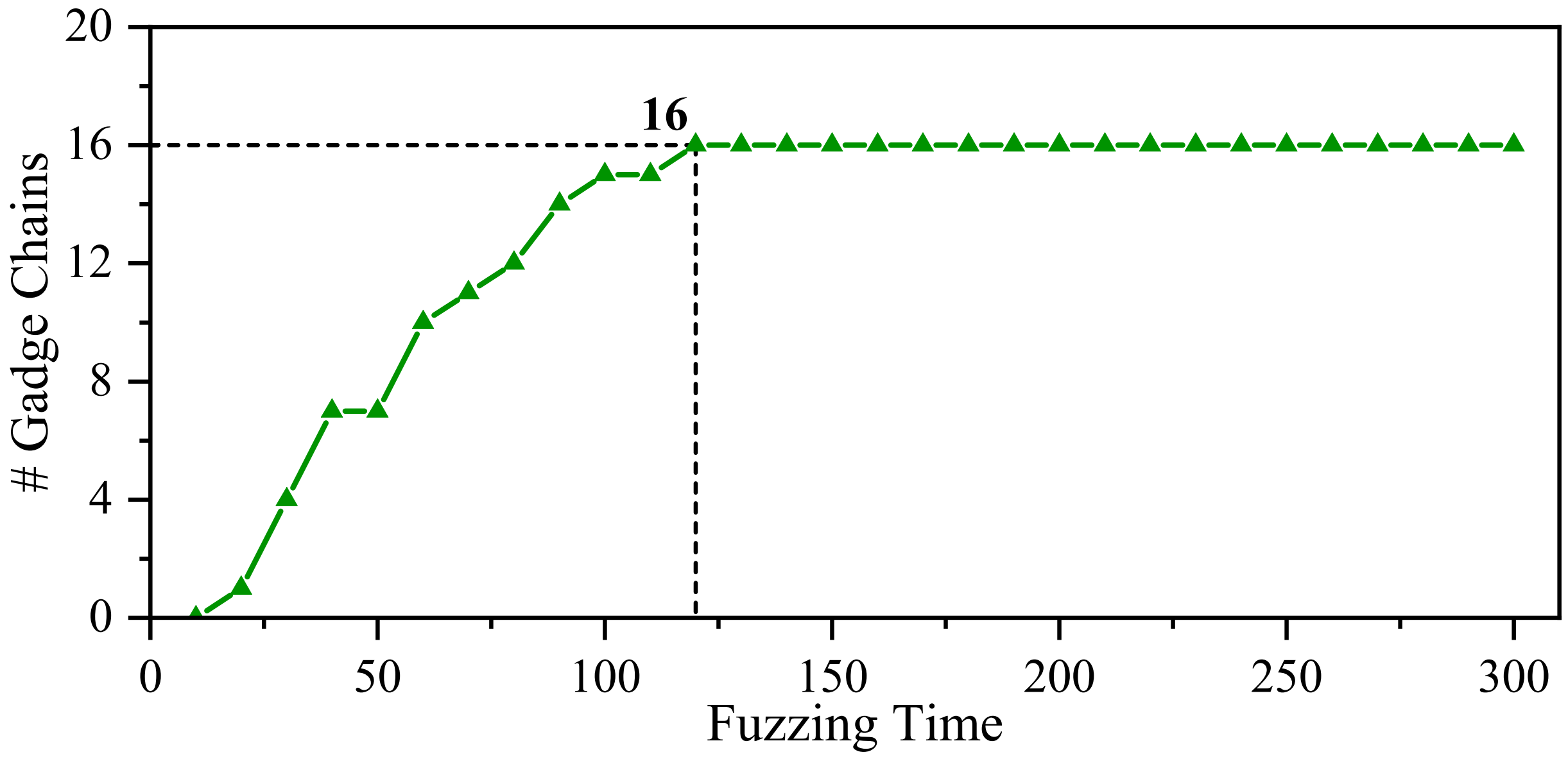}
\caption{\label{Timeout}Fuzzing time budget.}
\end{subfigure}
\caption{\label{parameter}Sensitivity analysis on gadget chain length and fuzzing time budget.}
\end{figure}

\noindent\textbf{Gadget Chain Length.}
Gadget chain length denotes the maximum quantity of gadgets that \sysname could chain in our static taint analysis. It is one of the critical hyperparameters in our approach because it determines the recall of gadget chain identification. For evaluation, we singly run the static identifier module of \sysname on ysoserial and counted the number of identified gadget chains within a specified maximum length, from 1 to 20 in increment of 1. For example, if the maximum threshold was set to 2, we would count the number of gadget chains with length of 1 and 2. According to the assessment of our employed security experts, it is reasonable to set the value range as $[1, 20]$ because the length of most publicly disclosed gadget chains is less than 20.

The evaluation results are shown in Figure \ref{Length}. We can find that the growth rate of the number of newly discovered gadget chains slows down (less than 10 chains for the first time) when the maximum gadget chain length is raised from 15 to 16. Therefore, we set the maximum gadget chain length to be 15 in Section \ref{Evaluation}.

\noindent\textbf{Fuzzing Time Budget.}
Fuzzing time budget represents the maximum time budget assigned to the fuzzer for validating a candidate gadget chain. It is another critical hyperparameter in our approach because assigning too much time to fuzz a gadget chain unable to be exploited will waste computation resources. Similar to the aforementioned experiment involving maximum gadget chain length, we used the 34 known gadget chains in ysoserial for evaluation. For each gadget chain, we ran the fuzzer 10 times and counted the number of validated gadget chains within a specified maximum length, from 0 to 300 seconds in increments of 10 seconds \cite{FUGIO}. As reported in Figure \ref{Timeout}, \sysname cannot successfully validate more gadgets after 120 seconds (2 minutes). Hence, we set the fuzzing time budget to be 120 seconds in all experiments in Section \ref{Evaluation}.

\begin{figure}[ht]
  \centering
  \includegraphics[width=\linewidth]{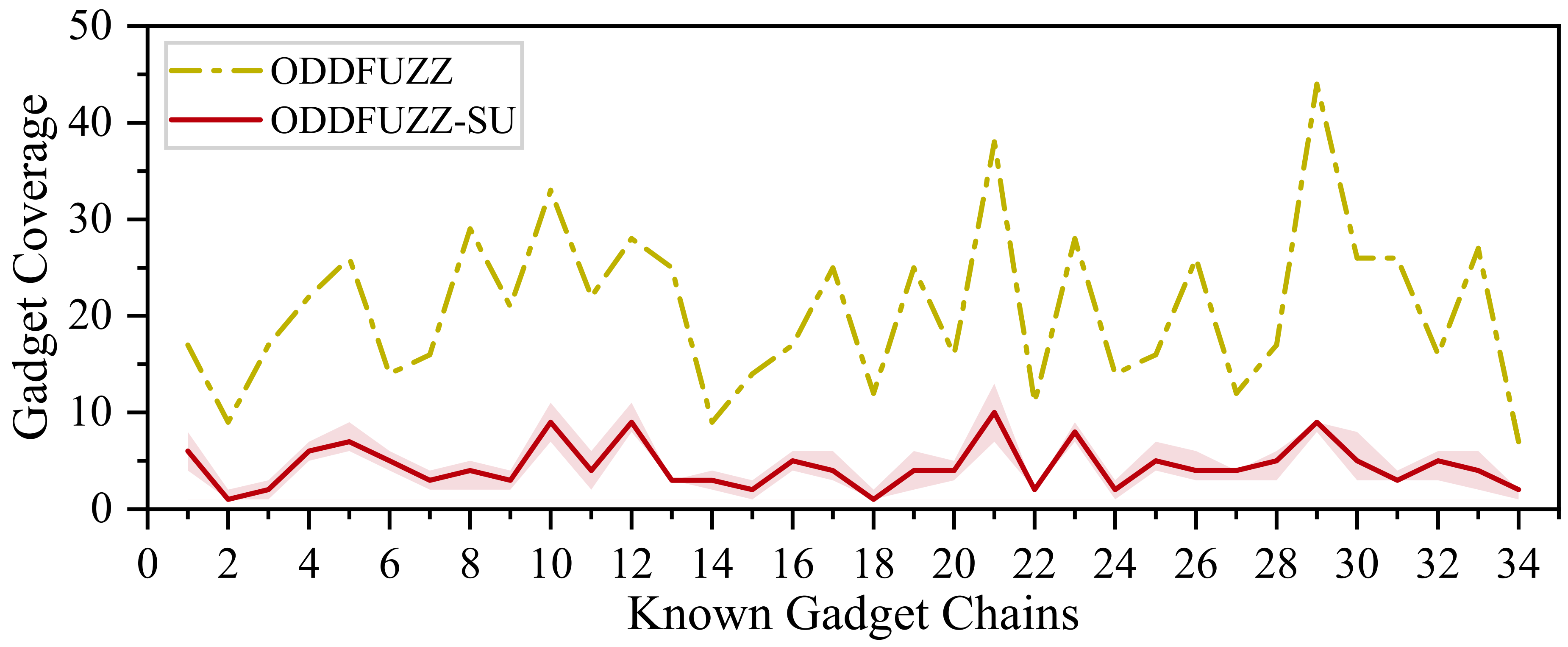}
\caption{Gadget coverage comparison of \sysname and \sysname-RM. The $x$-axis is 34 known gadget chains listed in Table \ref{34Chains} and the $y$-axis is the gadget coverage.}
\label{ConcreteCoverage1}
\end{figure}

\begin{figure}[t]
\centering
\begin{subfigure}{\linewidth}
\centering
\includegraphics[width=\linewidth]{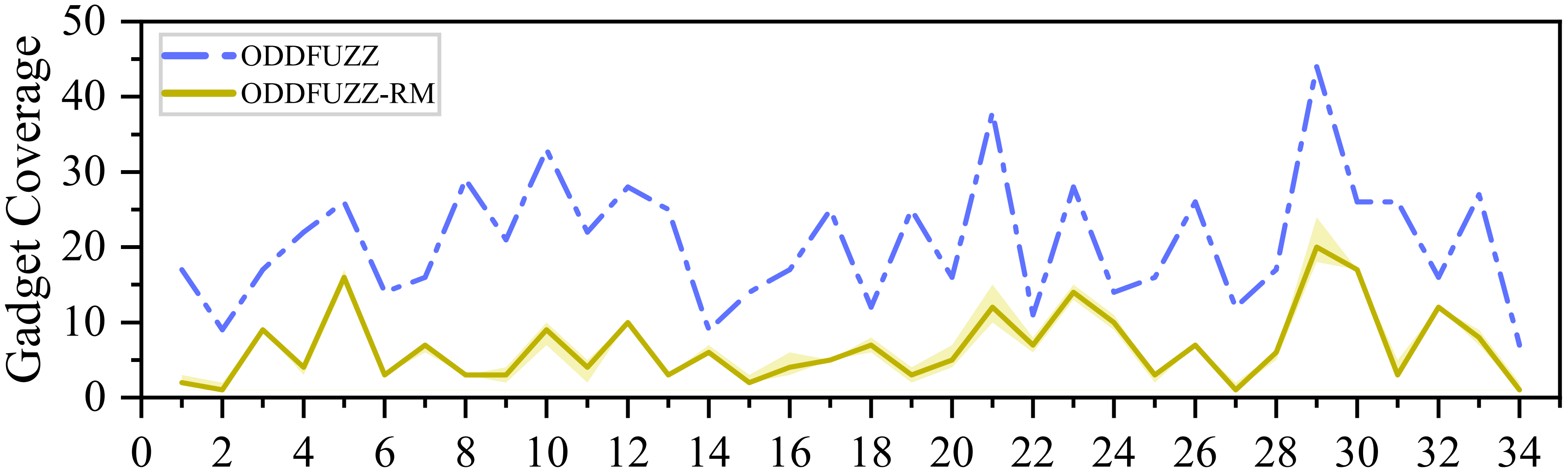}
\end{subfigure}
\begin{subfigure}{\linewidth}
\centering
\includegraphics[width=\linewidth]{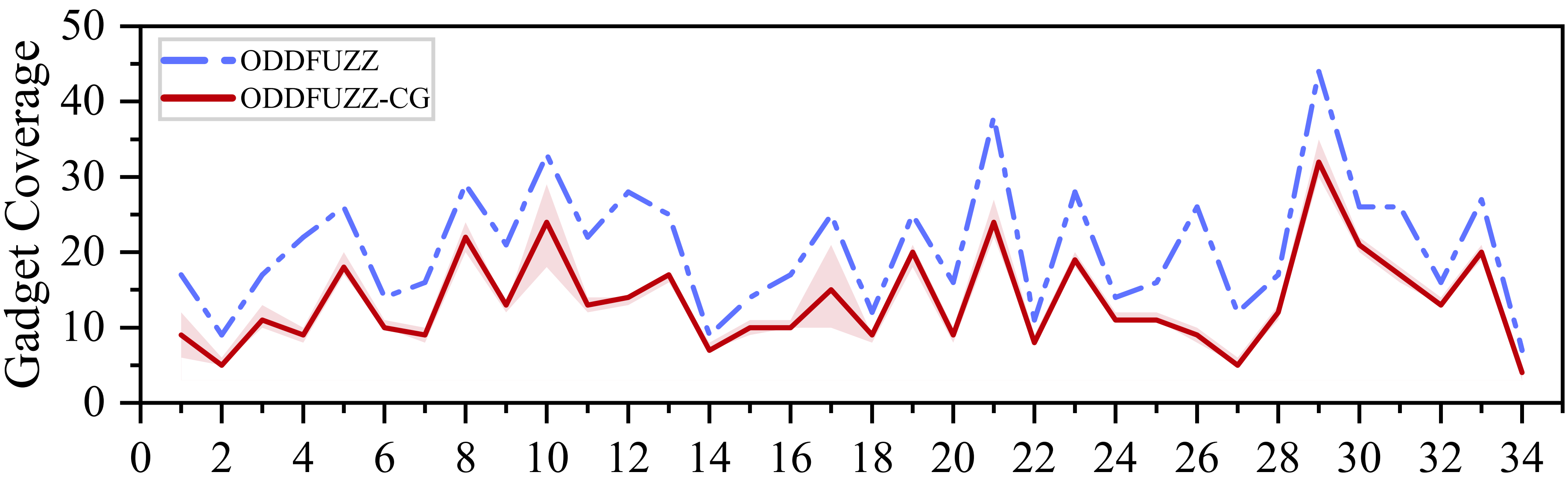}
\end{subfigure}
\begin{subfigure}{\linewidth}
\centering
\includegraphics[width=\linewidth]{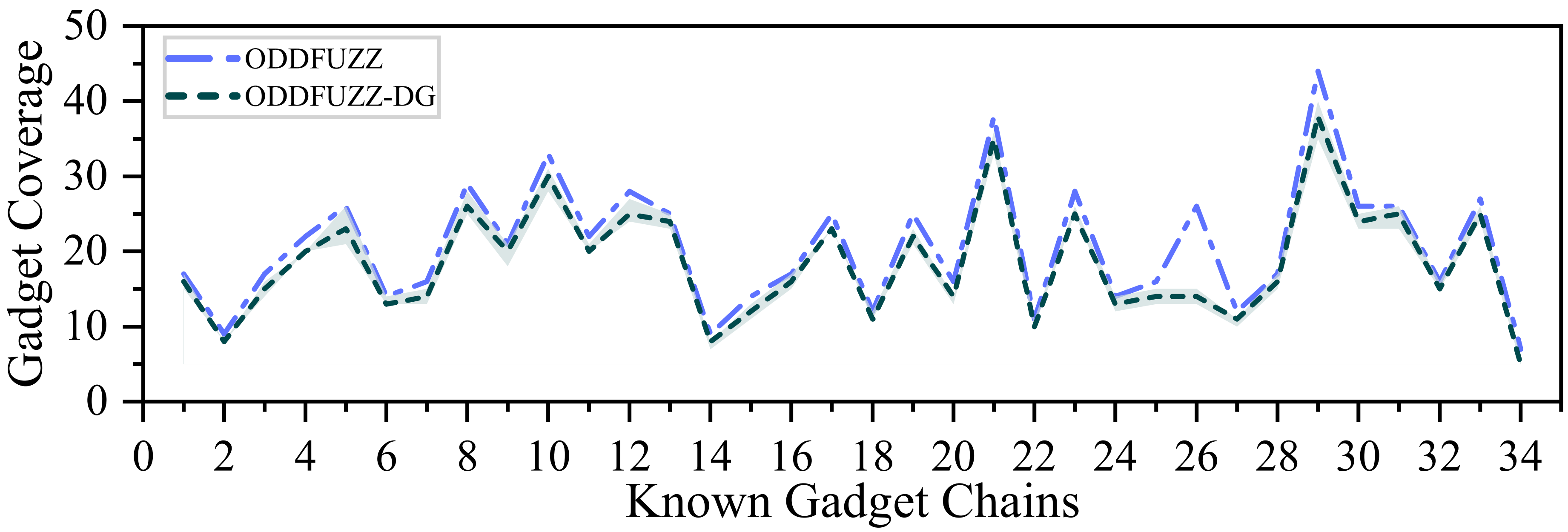}
\end{subfigure}
\caption{Gadget coverage comparison of \sysname, \sysname-RM, \sysname-CG, and \sysname-DG. The $x$-axis is 34 known gadget chains listed in Table \ref{34Chains}. The $y$-axis is the gadget coverage.}
\label{ConcreteCoverage2}
\end{figure}

\begin{table*}\scriptsize
 \caption{Evaluation results of \sysname on each gadget chain from ysoserial. \textbf{\# Gadgets} denotes the number of gadgets in each gadget chain. \code{Identified} and \code{Validated} respectively represents whether the gadget chain can be statically identified and dynamically validated.}
  \centering
  \renewcommand\arraystretch{1.2}
  \renewcommand\tabcolsep{3.4pt}
  \begin{tabular}{lll|c|cc|c|c}
    \toprule
    \multirow{2}*{\textbf{ID}} & \multirow{2}*{\textbf{Gadget Chain}} & \multirow{2}*{\textbf{Affected Version}} & \multirow{2}*{\textbf{\# Gadgets}} & \multicolumn{2}{c|}{\textbf{\sysname}} & \multirow{2}*{\textbf{GadgetInspector}} & \multirow{2}*{\textbf{SerHybrid}}\\
    ~ & ~ & ~ & ~ & Identified & Validated & ~ & ~ \\
    \midrule
    1  & \emph{AspectJWeaver} & aspectjweaver-1.9.2 & 9 & \Checkmark & \usym{2717} & - & - \\
    2  & \emph{BeanShell1} & bsh-2.0b5 & 6 & - & - & - & - \\ 
    3  & \emph{C3P0} & c3p0-0.9.5.2 & 6 & \Checkmark & \Checkmark & - & - \\ 
    4  & \emph{Click1} & click-nodeps-2.3.0 & 10 & \Checkmark & \Checkmark & - & - \\ 
    5  & \emph{Clojure} & clojure-1.8.0 & 10 & \Checkmark & \Checkmark & \Checkmark & - \\ 
    6  & \emph{CommonsBeanutils1} & commons-beanutils-1.9.2 & 5 & \Checkmark & \Checkmark & - & -  \\ 
    7  & \emph{CommonsCollections1} & commons-collections-3.1 & 7 & \Checkmark & \usym{2717} & \Checkmark & - \\ 
    8  & \emph{CommonsCollections2} & commons-collections4-4.0 & 13 & \Checkmark & \Checkmark & - & \Checkmark\\
    9  & \emph{CommonsCollections3} & commons-collections-3.1 & 13 & \Checkmark & \usym{2717} & - & -\\ 
    10 & \emph{CommonsCollections4} & commons-collections4-4.0 & 15 & \Checkmark & \Checkmark & - & - \\ 
    11 & \emph{CommonsCollections5} & commons-collections-3.1 & 8 & \Checkmark & \Checkmark & - & - \\ 
    12 & \emph{CommonsCollections6} & commons-collections-3.1 & 10 & \Checkmark & \Checkmark & - & \Checkmark\\ 
    13 & \emph{CommonsCollections7} & commons-collections-3.1 & 9 & \Checkmark & \Checkmark & - & -\\ 
    14 & \emph{FileUpload1} & commons-fileupload-1.3.1 & 3 & -& - & - & - \\ 
    15 & \emph{Groovy1} & groovy-2.3.9 & 10 & -& - & - & - \\ 
    16 & \emph{Hibernate1} & hibernate-core-4.3.11.Final & 7 & \Checkmark & \Checkmark & - & - \\ 
    17 & \emph{Hibernate2} & hibernate-core-4.3.11.Final & 9 & \Checkmark & \Checkmark & - & - \\ 
    18 & \emph{JBossInterceptors1} & jboss-interceptor-core:2.0.0.Final & 5 & -& - & -  & - \\
    19 & \emph{JRMPClient} & JDK-1.7 & 13 & -& - & - & - \\ 
    20 & \emph{JRMPListener} & JDK-1.7 & 9 & -& - & - & - \\ 
    21 & \emph{JSON1} & json-lib:jar-jdk15:2.4 & 22 & -& - & - & - \\ 
    22 & \emph{JavassistWeld1} & javassist-3.12.1.GA & 5 & -& - & - & - \\ 
    23 & \emph{Jdk7u21} & JDK-1.7 & 11 & - & - & - & - \\ 
    24 & \emph{Jython1} & jython-standalone-2.5.2 & 5 & \Checkmark & \usym{2717} & \Checkmark & - \\ 
    25 & \emph{MozillaRhino1} & js-1.7R2 & 8 & \Checkmark & \Checkmark & - & - \\ 
    26 & \emph{MozillaRhino2} & js-1.7R2 & 12 & \Checkmark & \Checkmark & - & - \\
    27 & \emph{Myfaces1} & myfaces-impl-2.2.9 & 4 & -& - & - & - \\
    28 & \emph{Myfaces2} & myfaces-impl-2.2.9 & 6 & -& - & - & - \\
    29 & \emph{ROME} & rome-1.0 & 15 & \Checkmark & \Checkmark & - & - \\ 
    30 & \emph{Spring1} & spring-core:4.1.4.RELEASE & 11 & -& - & - &- \\ 
    31 & \emph{Spring2} & spring-core:4.1.4.RELEASE & 12 & -& - & - & - \\ 
    32 & \emph{URLDNS}  & JDK & 7 & \Checkmark & \Checkmark & - & -\\ 
    33 & \emph{Vaadin1} & vaadin-server-7.7.14 & 10 & \Checkmark & \Checkmark & - & - \\ 
    34 & \emph{Wicket1} & wicket-util-6.23.0 & 3 & -& - & - & - \\
    \bottomrule
  \end{tabular}
  \label{34Chains}
\end{table*}

\begin{table}[t]\footnotesize
 \caption{List of previously unknown deserialization vulnerabilities discovered by \sysname.}
  \centering
  \renewcommand\tabcolsep{3.5pt}
  \begin{tabular}{cllccl}
    \toprule
    \textbf{No.} & \textbf{Application} & \textbf{Version} & \textbf{Impact} & \textbf{Status} & \textbf{CVE-ID} \\
    \midrule
    1  & WebLogic         & 12.2.1.4.0	& RCE &  Patched   & CVE-2020-14756  \\ 
    2  & WebLogic         & 12.2.1.4.0  & RCE &  Patched   & CVE-2020-14825\\ 
    3  & WebLogic         & 12.2.1.4.0  & RCE &  Patched   & CVE-2021-2135   \\ 
    4  & Sonatype Nexus   & 3.25.0      & RCE &  Patched   & CVE-2020-15871 \\
    5  & Apache Dubbo     & 2.7.7       & RCE &  Patched   & CVE-2020-11995\\
    6  & ProtoStuff       & 1.8.0       & RCE &  Reported  &  \\
    \bottomrule
  \end{tabular}
  \label{hunting}
\end{table}

\end{document}